\RequirePackage[normalem]{ulem} %DIF PREAMBLE
\RequirePackage{color}\definecolor{RED}{rgb}{1,0,0}\definecolor{BLUE}{rgb}{0,0,1} %DIF PREAMBLE
\documentclass[useAMS,usenatbib]{mn2e}

\usepackage{graphicx}
\usepackage[]{natbib}
\usepackage[]{subfig}
\usepackage{amssymb}
\usepackage{amsmath}
\usepackage{ulem}
\usepackage{mathrsfs}

\newcommand\msun{ M$_\odot$}%
\newcommand\sfrd{{$\rho_{\rm SFR}$}}%
\newcommand\smd{$\rho_*$}%
\newcommand\Msch{$M_{\rm C}$}%
\newcommand\Phis{$\Phi^*$}
\newcommand\Phid{$\Phi^*_2$}
\newcommand\Phione{$\Phi^*_1$}
\newcommand\alphas{$\alpha$}
\newcommand\alphad{$\alpha_2$}
\newcommand\alphaone{$\alpha_1$}
\newcommand\betas{$\beta$}
\newcommand\gammas{$\gamma$}
\newcommand\bigbox{Ref-L100N1504}
\newcommand\smallbox{Ref-L025N0376}
\newcommand\highres{Recal-L025N0752}
\newcommand\highresref{Ref-L025N0752}
\newcommand\Schaye{S15}
\newcommand\Eagle{\hbox{{\sc Eagle}}}
\newcommand\eagle{\hbox{{\sc Eagle}}}

\newcommand\Anarchy{{\sc Anarchy}}
\newcommand\FoF{{\sc FoF}}
\newcommand\Subfind{{\sc Subfind}}
\newcommand\OWLS{{\sc OWLS}}
\newcommand\GIMIC{{\sc GIMIC}}
\newcommand\Illustrius{{\sc Illustris}}

\newcommand\Cloudy{{\sc Cloudy}}

\newcommand\Primus{{\sc primus}}
\newcommand\ZFOURGE{{\sc zfourge}}
\newcommand\UVISTA{{\sc UltraVISTA}}

\title[Galaxy evolution in the \Eagle\ simulations]{Evolution of galaxy stellar masses and star formation rates in the \eagle\ simulations}
\author[M. Furlong et al]{M. Furlong $^{1}$\thanks{E-mail: michelle.furlong@durham.ac.uk}, 
 R. G. Bower$^1$, T. Theuns$^{1, 2}$, J. Schaye$^3$, R. A. Crain$^3$, M. Schaller$^1$,
\newauthor
C. Dalla Vecchia$^{4, 5}$, C. S. Frenk$^1$, I. G. McCarthy$^6$, J. Helly$^1$, A. Jenkins$^1$, 
\newauthor
and Y. M. Rosas-Guevara$^{7, 8}$    \\
$^{1}$Institute for Computational Cosmology, Durham University, South Road, Durham, DH1 3LE \\
$^2$Department of Physics, University of Antwerp, Groenenborgerlaan 171, B-2020 Antwerp, Belgium \\
$^{3}$Leiden Observatory, Leiden University, P.O. Box 9513, 2300 RA Leiden, the Netherlands \\
$^4$Instituto de Astrofs\'ica de Canarias, C/ V\'ia L\'actea s/n,38205 La Laguna, Tenerife, Spain \\
$^5$Departamento de Astrofs\'ica, Universidad de La Laguna, Av. del Astrofs\'icasico Franciso S\'anchez s/n, 38206 La Laguna, Tenerife, Spain \\
$^{6}$Astrophysics Research Institute, Liverpool John Moores University, 146 Brownlow Hill, Liverpool L3 5RF  \\
$^{7}$ Universit\'e de Lyon, Lyon, F-69003, France \\
$^{8}$ CNRS, UMR 5574, Centre de Recherche Astrophysique de Lyon, Ecole Normale Sup\'erieure de Lyon, Lyon, F-69007, France \\
}

\begin{document}

\date{Accepted 15 April 2015}

\pagerange{\pageref{firstpage}--\pageref{lastpage}} \pubyear{2015}

\maketitle

\label{firstpage}

\begin{abstract}
We investigate the evolution of galaxy masses and star formation rates
in the Evolution and Assembly of Galaxies and their Environment (\Eagle)
simulations. These comprise a suite of hydrodynamical simulations in a
$\Lambda$CDM cosmogony with subgrid models for radiative cooling, star
formation, stellar mass loss, and feedback from stars and accreting
black holes. The subgrid feedback was calibrated to reproduce the
observed present-day galaxy stellar mass function and galaxy sizes. Here
we demonstrate that the simulations reproduce the observed growth of the
stellar mass density to within 20 per cent. The simulation also tracks
the observed evolution of the galaxy stellar mass function out to
redshift $z=7$, with differences comparable to the plausible uncertainties
in the interpretation of the data. 
Just as with observed galaxies, the
specific star formation rates of simulated galaxies are bimodal, with distinct
star forming and passive sequences. 
The specific star formation rates
of star forming galaxies are typically 0.2 to 0.5 dex lower than
observed, but the evolution of the rates track the observations
closely. The unprecedented level of agreement between simulation and
data across cosmic time makes \Eagle\ a powerful resource to understand the physical
processes that govern galaxy formation.

\end{abstract}


\section{Introduction}

Although the basic model for how galaxies form within the framework of a cold dark matter cosmogony has been established for
many years \citep[e.g.][]{White78, White91}, many crucial aspects are still poorly understood. For example, what physical processes determine galaxy stellar masses and galaxy sizes? How do these properties evolve throughout cosmic history? How do stars and AGN regulate the evolution of galaxy properties?
Numerical simulations and theoretical models are a valuable tool for exploring these questions, but the huge dynamic range involved, and the complexity of the plausible underlying physics, limits the {\it ab initio} predictive power of such calculations \citep[e.g.][]{Schaye10,Aquilla}.

We recently presented the \eagle\ simulation project \citep[hereafter \Schaye]{Schaye14}, a suite of cosmological hydrodynamical simulations in which subgrid models parameterise our inability to faithfully compute the physics of galaxy formation below the resolution of the calculations. Calibrating the parameters entering the subgrid model for feedback by observations of the present-day galaxy stellar mass function (GSMF) and galaxy sizes, we showed that \eagle\ also reproduces many other properties of observed galaxies at $z \sim 0$ to unprecedented levels. The focus of this paper is to explore whether the good agreement, specifically that between the simulated and observed stellar masses and star formation rates, extends to higher redshifts.

Compared with semi-analytic models, hydrodynamical simulations such as \eagle\ have fewer degrees of freedom and have to make fewer simplifying assumptions to model gas accretion and the crucial aspects of the feedback from star formation and accreting black holes that is thought to regulate galaxy formation. They also allow the study of properties of the circumgalactic and intergalactic media, providing important complementary tests of the realism of the simulation. Such a holistic approach is necessary to uncover possible degeneracies and inconsistencies in the model. Having a calibrated and well-tested subgrid model is
of crucial importance, since it remains the dominant uncertainty in current simulations \citep{Aquilla}.

\Schaye\ present and motivate the subgrid physics implemented in \eagle. An overriding consideration of the parameterisation is that subgrid physics should only depend on local properties of the gas (e.g. density, metallicity), in contrast to other implementations used in the literature which for example depend explicitly on redshift, or on properties of the dark matter. Nevertheless, a physically reasonable set of parameters of the subgrid model for feedback exists for which the redshift $z\sim 0$ GSMF and galaxy sizes agree to within 0.2~dex with the observations. This level of agreement is unprecedented, and similar to the systematic uncertainty in deriving galaxy stellar masses from broad-band observations. Other observations of the local Universe, such as the Tully-Fisher relation, the mass-metallicity relation and the column density distribution functions of intergalactic CIV and OVI are also reproduced, even though they were not used in calibrating the model and hence could be considered \lq predictions\rq. 

In this paper we focus on the build-up of the stellar mass density, and the evolution of galaxy stellar masses and star formation rates, expanding the analysis of S15 beyond z$\sim 0$.  
A similar analysis was presented by \cite{Genel14}, for the \Illustrius\ simulation \citep{Illustrius}. 
They conclude that \Illustrius\ reproduces the observed evolution of the GSMF from redshifts 0 to 7 well, but we note that they used the star formation history in their calibration process.  
Another difference with respect to \cite{Genel14} is that we compare with recent galaxy surveys, which have dramatically tightened observational constraints on these measures of galaxy evolution.
For example 
\Primus\ \citep{Moustakas13}, \UVISTA\ \citep{Ilbert13, Muzzin13} and \ZFOURGE\ \citep{Tomczak13} provide improved constraints out to redshift 4. UV observations extend the comparison to even higher redshift, with inferred GSMFs available up to redshift 7 \citep{Gonzalez11, Duncan14}.
Observations of star formation rates also span the redshift range 0 to 7, with many different tracers of star formation (e.g.\ IR, radio, UV) providing consistency checks between data sets. 

This paper is organised as follows: In Section \ref{sec:sim} we provide a brief summary of \eagle\, in particular the subgrid physics used. In Section \ref{sec:sm} we compare the evolution of the stellar mass growth in the simulation to data out to redshift 7. We follow this with an analysis of the star formation rate density and specific star formation rates in Section \ref{sec:sfr}. In Section \ref{sec:discussion} we discuss the results and we summarise in Section \ref{sec:summary}. We generally find that the properties of the simulated galaxies agree with the observations to the level of the observational systematic uncertainties across all redshifts.

The \eagle\ simulation suite adopts a flat $\Lambda$CDM cosmogony with parameters from Planck \citep{Planck13};  $\Omega_\Lambda = 0.693$, $\Omega_m = 0.307$, $\Omega_b = 0.048$, $\sigma_8 = 0.8288$, $n_s = 0.9611$ and $H_0 = 67.77$ km s$^{-1}$ Mpc$^{-1}$. The \cite{Chabrier03} stellar initial mass function (IMF) is assumed in the simulations. 
Where necessary observational stellar masses and star formation rate densities have been renormalised to the Chabrier IMF\footnote {Specific star formation rates are not renormalised as the correction for star formation rates and stellar masses are similar and cancel each other.} and volumes have been rescaled to the Planck cosmology. 
Galaxy stellar masses are computed within a spherical aperture of 30 proper kiloparsecs (pkpc) from the centre of potential of the galaxy. 
This definition mimics a 2D Petrosian mass often used in observations, as shown in \Schaye. Star formation rates are computed within the same aperture. Distances and volumes are quoted in comoving units (e.g. comoving megaparsecs, cMpc), unless stated otherwise. Note that, unless explicitly stated, values are not given in $h^{-1}$ units.

\section{Simulations}
\label{sec:sim}
The \Eagle\ simulation suite consists of a large number of cosmological simulations, with variations that include parameter changes relative to those of the reference subgrid formulation, other subgrid implementations, different numerical resolutions, and a range of box sizes up to 100 cMpc boxes \citep[\Schaye, ][]{Crain15}. Simulations are denoted as, for example, L0100N1504, which corresponds to a simulation volume of L$=100$ cMpc on a side, using $1504^3$ particles of dark matter and an equal number of baryonic particles. A prefix distinguishes subgrid variations, for example Ref-L100N1504 is our reference model. 
These simulations use advanced smoothed particle hydrodynamics (SPH) and  state-of-the-art subgrid models to capture the unresolved physics.  
Cooling, metal enrichment, energy input from stellar feedback, black hole growth and feedback from AGN are included.  
The free parameters for stellar and AGN feedback contain considerable uncertainty (see \Schaye), and so are calibrated to the redshift 0.1 GSMF, with consideration given to galaxy sizes. A complete description of the code, subgrid physics and parameters can be found in \Schaye, while the motivation is given in \Schaye\ and \cite{Crain15}. Here we present a brief overview.

CAMB \citep[version Jan\_12]{CAMB} was used to generate the transfer function for the linear matter power spectrum with a Plank 1 \citep{Planck13} cosmology.
The Gaussian initial conditions were generated using the linear matter power spectrum and the random phases were taken from the public multi-scale white noise Panphasia field \citep{Jenkins13}. 
Particle displacements and velocities are produced at redshift 127 using second-order Langrangian perturbation theory \citep{Jenkins10}.
See Appendix B of \Schaye\ for more detail.

The initial density field is evolved in time using an extensively modified version of the parallel N-body SPH code Gadget-3 \citep{Springel08}, which is essentially a more computationally efficient version of the public code Gadget-2 described in detail by \cite{Springel05}.
In this Lagrangian code, a fluid is represented by a discrete set of particles, from which the gravitational and hydrodynamic forces are calculated.  
SPH properties, such as the density and pressure gradients, are computed by interpolating across neighbouring particles.

The code is modified to include updates to the hydrodynamics, as described in Dalla Vecchia et al. (in prep., see also \Schaye\ Appendix A), collectively referred to as \Anarchy. 
The impact of these changes on cosmological simulations are discussed in Schaller et al. (in prep.).
\Anarchy\ includes: 
\begin{itemize}
 \item The pressure-entropy formulation of SPH described in \cite{Hopkins13}. 
 \item The artificial viscosity switch of \cite{Cullen10} and an artificial conduction switch described by \cite{Price08}.
 \item A C2 \cite{Wendland95} kernel with 58 neighbours to interpolate SPH properties across neighbouring particles %to prevent {\bf ZZZ clumping ZZZ}.  
 \item The time step limiter from \cite{Durier12} that ensures feedback events are accurately modelled.
\end{itemize}

Two of the \Eagle\ simulations are analysed in this paper\footnote{Two further simulations are considered in Appendix \ref{ap:res}.}.
The first \Eagle\ simulation analysed in this paper is \bigbox, a ($100$ cMpc)$^3$ periodic box with $2 \times 1504^3$ particles.  Initial masses for gas particles are $1.81 \times 10^6$\msun\ and masses of dark matter particles are $9.70 \times 10^6$\msun. 
Plummer equivalent comoving gravitational softenings are set to 1/25 of the initial mean inter-particle spacing and are limited to a maximum physical size of $0.70$ pkpc.

We also use simulation \highres\ which has 8 times better mass resolution and 2 times better spatial resolution in a (25 cMpc)$^3$ box.
The box sizes, particle numbers and resolutions are summarised in Table \ref{table:boxes}.
Note that subgrid stellar feedback parameters and black hole growth and feedback parameters are recalibrated in the \highres\ simulation, as explained in Section \ref{sec:simres}. 

\begin{table*}
\centering
\caption{Box size, particle number, baryonic and dark matter particle mass, comoving and maximum proper gravitational softening for \bigbox\ and \highres\ simulations.}
\smallskip
 \begin{minipage}{8.5cm}
\centering
  \begin{tabular}{|l |c | c| c| c| c| c|}
   \hline
   Simulation &  L & N & $m_{\rm g}$ & $m_{\rm dm}$ & $\epsilon_{\rm com}$ & $\epsilon_{\rm prop}$ \\
   \hline
    & [cMpc] & & [M$_\odot$] & [M$_\odot$] & [ckpc] & [pkpc] \\
   \hline
   Ref-L100N1504  & 100  & $2\times (1504)^3$ & 1.81$\times 10^6$ & 9.70$\times 10^6$ & 2.66 & 0.70 \\
   Recal-L025N0752  &  25  &  $2\times (752)^3$ & 2.26$\times 10^5$ & 1.21$\times 10^6$ & 1.33 & 0.35 \\
   \hline
  \end{tabular}  \par 
  \end{minipage}
\label{table:boxes}
\end{table*} 

\subsection{Subgrid physics}
The baryonic subgrid physics included in these simulations is broadly based on that used for the \OWLS\ \citep{Schaye10} and \GIMIC\ \citep{Crain09}
projects, although many improvements, in particular to the stellar feedback scheme and black hole growth, have been implemented. 
We emphasise that all subgrid physics models depend solely on local inter-stellar medium (ISM) properties. 

\begin{itemize}
 \item \underline{Radiative cooling and photo-heating} in the simulation are included as in \cite{Wiersma09a}.  
The element-by-element radiative rates are computed in the presence of the cosmic microwave background (CMB) and the \cite{HaardtMadau01} model for UV and X-ray background radiation from quasars and galaxies.  
The eleven elements that dominate radiative cooling are tracked, namely H, He, C, N, O, Ne, Mg, Si, Fe, Ca and Si. The cooling tables, as a function of density, temperature and redshift are produced using \Cloudy, version 07.02 \citep{Cloudy}, assuming the gas is optically thin and in photoionization equilibrium.  

Above the redshift of reionization the CMB and a \cite{HaardtMadau01} UV-background up to 1 Ryd, to account for photo-dissociation of H$_2$, are applied. Hydrogen reionization is implemented by switching on the full \cite{HaardtMadau01} background at redshift $11.5$.

 \item \underline{Star formation} is implemented following \cite{Schaye08}.  
Gas particles above a metallicity-dependent density threshold, n$^*_{\rm H}$(Z), have a probability of forming stars, determined by their pressure.  
The Kennicutt-Schmidt star formation law \citep{Kennicutt98}, under the assumption of disks in vertical hydro-static equilibrium, can be written as 
\begin{equation}
\dot{m}_* = m_{\rm g}A(1{\rm M}_\odot {\rm pc}^{-2})^{-n} (\frac{\gamma}{G} f_{\rm g} P)^{(n-1)/2},
\end{equation}
where $m_{\rm g}$ is the gas particle mass, $A$ and $n$ are the normalisation and power index of the Kennicutt-Schmidt star formation law, $\gamma$ = 5/3 is the ratio of specific heats, G is the gravitational constant, $f_g$ = 1 is the gas fraction of the particle and P is its pressure.
As a result the imposed star formation law is specified by the observational values of $A=1.515 \times 10^{-4}$ M$_\odot$yr$^{-1}$kpc$^{-2}$ and $n=1.4$, where we have decreased the amplitude by a factor of 1.65 relative to the value of \cite{Kennicutt98} to account for the use of a Chabrier, instead of Salpeter, IMF.

As we do not resolve the cold gas phase, a star formation threshold above which cold gas is expected to form is imposed.
The star formation threshold is metallicity dependent and given by 
\begin{equation}
n^*_{\rm H}(Z) = 0.1{\rm cm}^{-3} \left(\frac{Z}{0.002}\right)^{-0.64},
\end{equation}
where Z is the metallicity \citep[from][, eq 19 and 24, also used in SFTHRESHZ model of the OWLS project]{Schaye04}.

A pressure floor as a function of density is imposed, of the form $P \propto \rho^{\gamma_{\rm eff}}$, for gas with density above $n^*_{\rm H}(Z)$ and $\gamma_{\rm eff} = 4/3$. This models the unresolved multi-phase ISM. Our choice for $\gamma_{\rm eff}$ ensures that the Jeans mass is independent of density and prevents spurious fragmentation provided the Jeans mass is resolved at $n^*_{\rm  H}(Z)$ \citep[see][]{Schaye08}.
Gas particles selected for star formation are converted to collisionless star particles, which represent a simple stellar population with a \cite{Chabrier03} IMF.

 \item \underline{Stellar evolution and enrichment} is based on \cite{Wiersma09b} and detailed in \Schaye.  
Metal enrichment due to mass loss from AGB stars, winds from massive stars, core collapse supernovae and type Ia supernovae of the 11 elements that are important for radiative cooling are tracked, using the yield tables of \cite{Marigo01}, \cite{Portinari98} and \cite{Thielemann03}.
The total and metal mass lost from stars are added to the gas particles that are within an SPH kernel of the star particle.

 \item \underline{Stellar feedback} is treated stochastically, using the thermal injection method described in \cite{DallaVecchia12}.  
The total available energy from core collapse supernovae for a Chabrier IMF assumes all stars in the stellar mass range $6$$-100$\msun\footnote{6 - 8\msun\ stars explode as electron capture supernovae in models with convective overshoot, e.g. \cite{Chiosi92}.} release $10^{51}$ erg of energy into the ISM and the energy is injected  after a delay of $30$ Myr from the time the star particle is formed.
Rather than heating all gas particle neighbours within the SPH kernel, neighbours are selected stochastically based on the available energy, then heated by a fixed temperature difference of $\Delta \rm T = 10^{7.5}$K.
The stochastic heating distributes the energy over less mass than heating all neighbours.
This results in a longer cooling time relative to the sound crossing time across a resolution element,  allowing the thermal energy to be converted to kinetic energy, thereby limiting spurious losses \citep{DallaVecchia12}.

In \Eagle, the fraction of this available energy injected into the ISM depends on the local gas metallicity and density.  
The stellar feedback fraction, in units of the available core collapse supernova energy, is specified by a sigmoid function,
\begin{equation}
f_{\rm th} = f_{\rm th, \rm min} + \frac{f_{\rm th, \rm max} - f_{\rm th, \rm min}}{1 + \left( \frac{Z}{0.1Z_\odot} \right)^{n_Z} \left( \frac{n_{H, \rm birth}}{n_{\rm H, 0}} \right)^{-n_{\rm n}}},
\end{equation}
where $Z$ is the metallicity of the star particle, $n_{\rm H, \rm birth}$ is the density of the star particle's parent gas particle when the star was formed and $Z_\odot = 0.0127$ is the solar metallicity.

The values for $f_{\rm th, \rm max}$ and $f_{\rm th, \rm min}$, the parameters for the maximum and minimum energy fractions, are fixed at 3 and 0.3 for both simulations analysed here.
At low Z and high $n_{\rm H, \rm birth}$, $f_{\rm th}$ asymptotes towards $f_{\rm th, \rm max}$  and at high Z and low $n_{\rm H, \rm birth}$ asymptotes towards $f_{\rm th, \rm min}$.
Applying up to 3 times the available energy can be justified by appealing to the different forms of stellar feedback, e.g. supernova, radiation pressure, stellar winds which are not treated separately here as we do not have the resolution to resolve these forms of stellar feedback.
This also offsets the remaining numerical radiative losses \citep{Crain15}.

The power law indexes are $n_{\rm Z} = n_{\rm n} = 2/\ln(10)$ for the Ref model, with $n_{\rm n}$ changed to $1/\ln(10)$ for the Recal model, resulting in weaker dependence of $f_{\rm th}$ on the density in the high resolution model.
The normalisation of the density term, $n_{\rm H, 0}$, is set to 0.67 cm$^{-3}$ for the Ref model and to 0.25 cm$^{-3}$ for the Recal model. The feedback dependence is motivated in \cite{Crain15}.

 \item \underline{Black hole seeding and growth} is implemented as follows.
Halos with a mass greater than $10^{10}$ h$^{-1}$M$_\odot$ are seeded with a black hole of $10^5$ h$^{-1}$M$_\odot$, using the method of \cite{Springel05b}.  
Black holes can grow through mergers and accretion.
Accretion of ambient gas onto black holes follows a modified Bondi-Hoyle formula that accounts for the angular momentum of the accreting gas \citep{Rosas-Guevara14}.
Differing from, e.g. \cite{Springel05b}, \cite{Booth&Schaye09}, \cite{Rosas-Guevara14}, the black hole accretion rate is not increased relative to the standard Bondi accretion rate in high-density regions.

For the black hole growth there is one free parameter, $C_{\rm visc}$, which is used to determine the accretion rate from
\begin{equation}
\dot{m}_{\rm accr} = {\rm min}(\dot{m}_{\rm bondi}\left[ C^{-1}_{\rm visc}({\rm c}_{\rm s}/\rm V_\Phi)^3 \right], \dot{m}_{\rm bondi}),
\end{equation}
where c$_{\rm s}$ is the sound speed and V$_\Phi$ is the rotation speed of the gas around the black hole. The Bondi rate is given by
\begin{equation}
\dot{m}_{\rm bondi} = \frac{4\pi G^2 m_{\rm BH}^2 \rho}{({\rm c}_{\rm s}^2+{\rm v}^2)^{3/2}},
\end{equation}
where v is the relative velocity of the black hole and the gas.
The accretion rate is not allowed to exceed the Eddington rate,  $\dot{m}_{\rm Edd}$, given by

\begin{equation}
\dot{m}_{\rm Edd} = \frac{4\pi G m_{\rm BH} m_{\rm p}}{\epsilon_{\rm r} \sigma_{\rm T} c},
\end{equation} 
where $m_{\rm p}$ is the proton mass, $\sigma_{\rm T}$ is the Thomson scattering cross section and $\epsilon_{\rm r}$ is the radiative efficiency of the accretion disc.
The free parameter $C_{\rm visc}$ relates to the viscosity of the (subgrid) accretion disc and $\left( {c_{\rm s}}/{\rm V_\Phi} \right)^3 / C_{\rm visc}$ relates the Bondi and viscous time scales \citep[see][for more detail]{Rosas-Guevara14}.

\item \underline{AGN feedback} follows the accretion of mass onto the black hole.  
A fraction of the accreted gas is released as thermal energy into the surrounding gas.  
Stochastic heating, similar to the supernova feedback scheme, is implemented with a fixed heating temperature $\Delta$T$_{\rm AGN}$, where $\Delta$T$_{\rm AGN}$ is a free parameter.
The method used is based on that of \cite{Booth&Schaye09} and \cite{DallaVecchia08}, see \Schaye\ for more motivation.
\end{itemize}

The effect of varying some of the subgrid parameters is explored in \cite{Crain15}.
The values of the parameters that differ between the two simulations used in this paper, \bigbox\ and \highres\, are listed in Table \ref{table:params}.

\subsection{Resolution Tests}
\label{sec:simres}
\begin{table}
\centering
\caption{Values of parameters that differ between \bigbox\ and \highres.}
\smallskip
 \begin{minipage}{8.5cm}
\centering
  \begin{tabular}{|l |c | c| c| l|}
   \hline
   {\bf Simulation Prefix} & $n_{\rm H, 0}$ & $n_{\rm n}$ & $C_{\rm visc}$\footnote{Note that the subgrid scheme is not very sensitive to the changes in $C_{\rm visc}$, as shown in Appendix B of \cite{Rosas-Guevara14}.} & $\Delta$T$_{\rm AGN}$\\
   \hline
    & [cm$^{-3}$] & & & [K] \\
   \hline
   Ref     & 0.67  & 2/$\ln$(10) & 2$\pi$ 				& 10$^{8.5}$\\
   Recal   & 0.25  & 1/$\ln$(10) & 2$\pi \times 10^3$   & 10$^{9}$\\
   \hline
  \end{tabular}  \par 
  \end{minipage}
\label{table:params}
\end{table}

We distinguish between the {\it strong} and {\it weak} numerical convergence of our simulations, as defined and motivated in \Schaye. By strong convergence we mean that simulations of different resolutions give numerically converged answer, {\it  without} any change to the subgrid parameters. 
In \Schaye\ it is argued that strong convergence is not expected from current simulations, as higher-resolution often implies changes in the subgrid models, for example energy injected by feedback events often scales directly with the mass of the star particle formed. In addition, with higher resolution the physical conditions of the ISM and hence the computed radiative losses, will change. Without turning off radiative cooling or the hydrodynamics (which could be sensitive to the point at which they are turned back on), the changes to the ISM and radiative losses are expected to limit the strong convergence of the simulation.

The \Eagle\ project instead focuses on demonstrating that the simulations shows good weak convergence (although \Schaye\ shows that the strong convergence of the simulation is on par with other hydrodynamical simulations). Weak convergence means that simulations of different resolutions give numerically converged results, {\it  after} recalibrating one or more of the subgrid parameters. 
As it is argued in \Schaye\ that current simulations cannot make {\it  ab initio} predictions for galaxy properties, due to the sensitivity of the results to the parameters of the subgrid models for feedback, and calibration is thus required, the high-resolution \Eagle\ simulation subgrid parameters are recalibrated to the same observable (the present-day GSMF, galaxy sizes, and the stellar-mass black hole mass correlation) as the standard resolution simulations.
This recalibrated high-resolution model, \highres, enables us to test the weak convergence behaviour of the simulation and to push our results for galaxy properties to 8 times lower stellar mass.
In Table \ref{table:params} we highlight the parameters that are varied between the Ref and Recal models.
In the main text of this paper we consider weak convergence tests, strong convergence tests can be found in Appendix \ref{ap:res}.

As a simulation with a factor of 8 better mass resolution requires a minimum of 8 times the CPU time (in practice the increase in time is longer due to the higher-density regions resulting in shorter time steps and difficulties in producing perfectly scalable algorithms), we compare the (100 cMpc)$^3$ intermediate-resolution simulation to a (25 cMpc)$^3$ high-resolution simulation.
Note that for volume averaged properties the (25 cMpc)$^3$ box differs from the (100 cMpc)$^3$ box not only due to the resolution but also due to the absence of larger objects and denser environments in the smaller volume.  
As a result, for volume averaged quantities we present only the \bigbox\ simulation in the following sections and revisit the convergence of these quantities in Appendix \ref{ap:res}.
For quantities as a function of stellar mass we present both the \bigbox\ and \highres\ simulations, although the comparison at high redshifts is limited by the small number of objects in the high-resolution simulation, which has a volume that is 64 times smaller.

\subsection{Halo and galaxy definition}
\label{sec:galdef}

Halo finding is carried out by applying the friends-of-friends (\FoF) method \citep{Davis85} on the dark matter, with a linking length of $0.2$ times the mean inter-particle separation.  Baryonic particles are assigned to the group of their nearest dark matter particle.
Self-bound overdensities within the group are found using \Subfind\ \citep{Springel01,Dolag09}; these substructures are the galaxies in our simulation.
A `central' galaxy is the substructure with the largest mass within a halo.  All other galaxies within a halo are `satellites'.
Note that any \FoF\ particles not associated with satellites are assigned to the central object, thus the mass of a central galaxy may extend throughout its halo.

A galaxy's stellar mass is defined as the stellar mass associated with the subhalo within a 3D 30 pkpc radius, centred on the minimum of the subhalo's centre of gravitational potential.  
Only mass that is bound to the subhalo is considered, thereby excluding mass from other subhalos.
This definition is equivalent to the total subhalo mass for low mass objects, but excludes diffuse mass around larger subhalos, which would contribute to the intra-cluster light (ICL).
\Schaye\ shows that this aperture yields results that are close to a 2D Petrosian aperture, often used in observations, e.g. \cite{LiandWhite09}.
The same 3D 30~pkpc aperture is applied when computing the star formation rates in galaxies, again considering only particles belonging to the subhalo.
The aperture constraint has only a minimal effect on the star formation rates because the vast majority of star formation occurs in the central 30 pkpc, even for massive galaxies.

\section{Evolution of galaxy stellar masses}
\label{sec:sm}

We will begin this section by comparing the growth in stellar mass density across cosmic time in the largest \Eagle\ simulation, \bigbox, to a number of observational data sets.
This is followed with a comparison of the evolution of the galaxy stellar mass function (GSMF) from redshift 0 to 7 and a discussion on the impact of stellar mass errors in the observations.  
We also consider the convergence of the GSMF in the simulation at different redshifts.

\subsection{The stellar mass density}
\label{sec:smdensity}

\begin{figure*}
  \centering
  \includegraphics[width=1.0\textwidth]{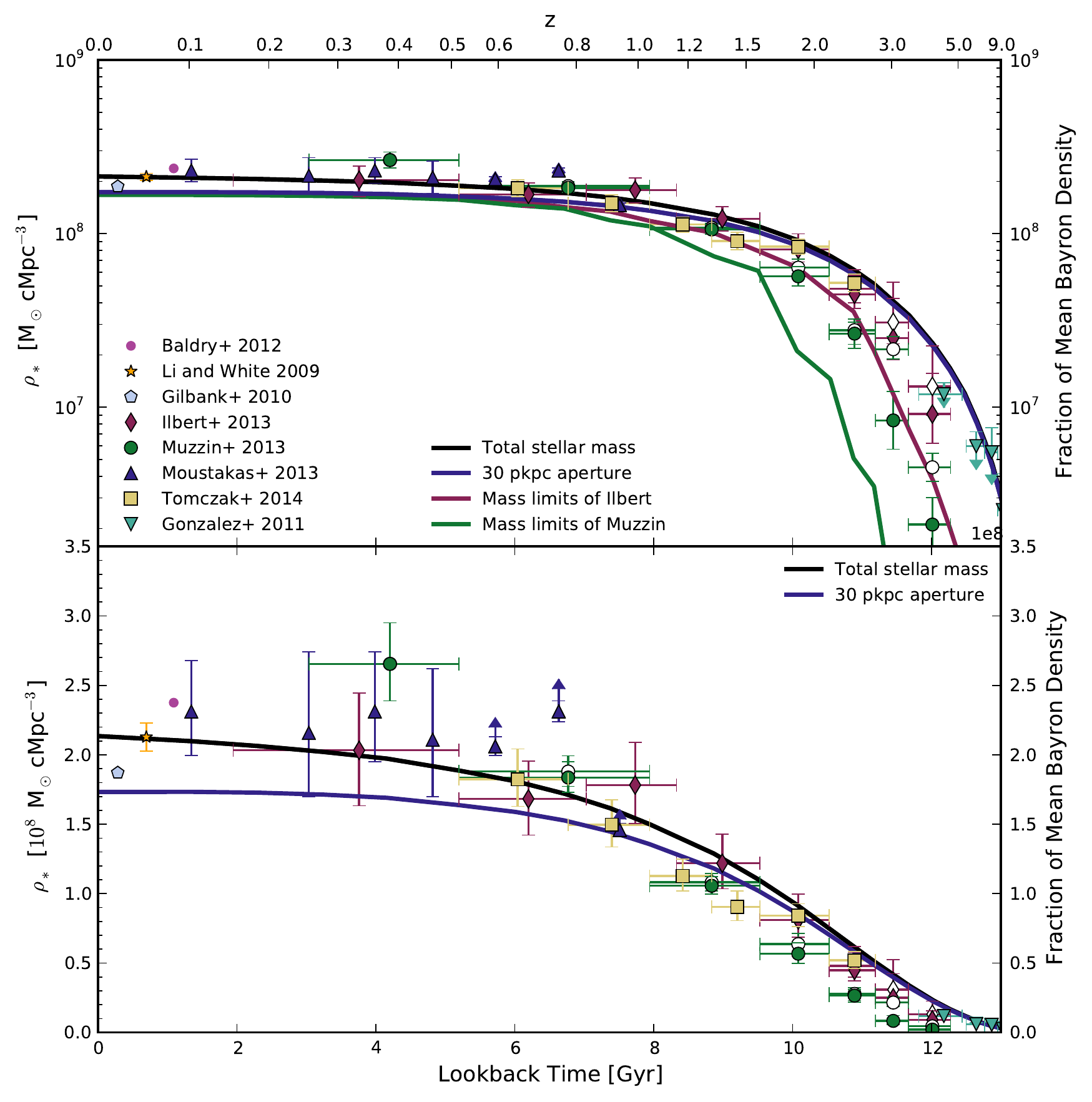}
  \caption{
	The stellar mass density as a function of time on a log and linear scale (top and bottom panels, respectively). 
	The black solid curve is the total stellar mass density from the \eagle\ simulation Ref-L0100N1504, and the blue curve is the stellar mass density in galaxies in that simulation (i.e. excluding intra-cluster light). 
	Observational data are plotted as symbols, see the legend for the original source.
	Open symbols refer to observations that include extrapolations of the GSMF below the mass completeness of the survey, filled symbols are the raw data.  
	Where necessary, data sets have been scaled to a Chabrier IMF and the Planck cosmology, as used in the simulation.
	The {\bf top panel} shows \smd\ for all galaxies in the simulation in {\it blue} and \smd\ for galaxies above the completeness limit of observations by \protect \cite{Ilbert13} and \protect \cite{Muzzin13} in red and green, respectively.  The corresponding data sets for \citet{Ilbert13} and \citet{Muzzin13} are coloured accordingly, and simulation lines should be compared to corresponding filled red and green symbols.	The {\bf bottom panel} shows \smd\ on a linear scale.
	From redshift 0 to 0.5, \smd\ in galaxies agrees with the observations at the 20\% level, with the simulated \smd\ lower by around 0.1~dex. At redshifts from 0.5 to 7, the model agrees well with the data, although the level of agreement above redshift 2 depends on the assumed incompleteness correction.
 }
  \label{fig:smdensity}
\end{figure*}

We begin the study of the evolution in the primary \Eagle\ simulation, \bigbox, by considering the build up of stellar mass. We present the stellar mass density (\smd) as a function of lookback time
in Figure \ref{fig:smdensity}, with redshift on the upper axis. Plotting the stellar mass density as a function of time (rather than redshift, say) gives a better visual impression of how much different epochs contribute to the net stellar build-up. 
 
We added to this figure recent observational estimates of \smd\ from a number of galaxy surveys.  
Around redshift 0.1 we show data from \cite{Baldry12} (GAMA survey), \cite{LiandWhite09} (SDSS), \cite{Gilbank10} (Stripe82 - SDSS) and \cite{Moustakas13} (\Primus).  
The values agree to within $0.55 \times 10^8$\msun cMpc$^{-3}$, which is better than 0.1 dex.
The \cite{Moustakas13} data set extends to redshift one, providing an estimate for \smd\ for galaxies with masses greater than $10^{9.5}$\msun.
Note, however, that above redshift 0.725 the \cite{Moustakas13} measurements of \smd\ are a lower limit as they only include galaxies with stellar masses of $10^{10}$\msun\ or above.
\cite{Ilbert13} and \cite{Muzzin13} estimate \smd\ from redshifts 0.2 to 4 from the \UVISTA\ survey.  
These two data sets use the same observations but apply different signal-to-noise limits and analyses to infer stellar masses resulting in slightly different results.
We include both studies in the figure to asses the intrinsic systematics in the interpretation of the data.
Both data sets extrapolate the observations to $10^8$\msun\ to estimate a \lq total\rq\ stellar mass density. The data sets are consistent within the estimated error bars up to redshift 3.
Above redshift 3 they differ, primarily because of the strong dependence of \smd\ on how the extrapolation below the mass completeness limit of the survey is performed.
The estimated \smd\ from observed galaxies can be compared to the extrapolated \smd\ for both data sets by comparing the filled and open symbols in Figure \ref{fig:smdensity}.
\cite{Tomczak13} estimate stellar mass densities between redshifts 0.5 and 2.5 from the \ZFOURGE\ survey.  The mass completeness limits for this survey are below $10^{9.5}$\msun\ at all redshifts, probing lower masses than other data sets at the same redshifts. For this data set no extrapolation is carried out in estimating \smd. In the simulations, galaxies with masses below $10^{9}$\msun\ contribute only 12\% to the stellar mass density at redshift 2 and their contribution decreases with decreasing redshift due to the flattening of the GSMF (see Section \ref{sec:gsmf}). 

At redshifts below two the various observational measurements show agreement on the total stellar mass density to better than 0.1 dex. From redshift 2 to 4 the agreement is poorer, with differences up to 0.4~dex, primarily as a result of applying different extrapolations to correct for incompleteness.
At redshifts above four only the UV observations of \cite{Gonzalez11} are shown.  Note that these do not include corrections for nebular emission lines and hence may overestimate \smd\ \citep[e.g.][]{Smit14}.
We therefore plot these values for \smd\ as upper limits.

\begin{table}
\centering
\caption{Mass completeness limit at redshifts 0.2 to 4 for GSMF observations of \protect \cite{Ilbert13} and \protect \cite{Muzzin13}.}
\smallskip
 \begin{minipage}{8.5cm}
\centering
  \begin{tabular}{|l |c | c|}
   \hline
   {\bf Redshift} & {\bf \protect \cite{Ilbert13}} & {\bf \protect \cite{Muzzin13}} \\
   \hline
    & {log$_{10}$(M$_*$) $\lbrack$M$_\odot \rbrack$} & {log$_{10}$(M$_*$) $\lbrack$M$_\odot \rbrack$} \\
   \hline
   0.2 - 0.5  & 7.93   &  8.37  \\
   0.5 - 0.8\footnote{\protect \cite{Muzzin13} use redshift ranges 0.5 to 1.0 and 1.0 to 1.5.}  & 8.70   &  8.92\\
   0.8 - 1.1  & 9.13   &  -\\
   1.1 - 1.5  & 9.42   &  9.48\\
   1.5 - 2.0  & 9.67   &  10.03\\
   2.0 - 2.5  & 10.04  &  10.54\\
   2.5 - 3.0  & 10.24  &  10.76\\
   3.0 - 4.0  & 10.27  &  10.94\\
   \hline
  \end{tabular}  \par 
   \vspace{-0.75\skip\footins}
   \renewcommand{\footnoterule}{}
  \end{minipage}
\label{table:obsmass}
\end{table}

The solid black line in each panel of Figure \ref{fig:smdensity} shows the build up of \smd\ in the simulation.
The log scale used in the upper panel emphasises the rapid fractional increase at high redshift.
There is a rapid growth in \smd\ from the early universe until 8 Gyr ago, around redshift 1, by which point 70$\%$ of the present day stellar mass has formed.  
The remaining 30$\%$ forms in the 8 Gyr, from redshift 1 to 0.  
We find that 50$\%$ of the present day stellar mass was in place 9.75 Gyr ago, by redshift 1.6. 

The simulation is in good agreement with the observed growth of stellar mass across the whole of cosmic time, falling within the error bars of the observational data sets.  
We find that 3.5\% of the baryons are in stars at redshift zero, which is close to the values of 3.5\% and 4\% reported by \cite{LiandWhite09} and \cite{Baldry12}, respectively. 

However, it should be noted that observed stellar mass densities are determined by integrating the GSMF, thereby excluding stellar mass associated with intra-cluster light (ICL).
To carry out a fairer comparison, we apply a 3D 30~pkpc aperture to the simulated galaxies to mimic a 2D Petrosian aperture, as applied to many observations (see Section \ref{sec:galdef} and \Schaye).  
The aperture masses more accurately represent the stellar light that can be detected in observations. 
The result of the aperture correction is shown as a solid blue line in both panels \footnote{Note the mass in the simulation associated with the ICL resides in the largest halos, as will be shown in a future paper.}.

In this more realistic comparison of the model to observations, which excludes the ICL, we find that from high redshift to redshift 2 there is little difference between the total  \smd\ and the aperture stellar mass density associated with galaxies.  
At these high redshifts the simulation curve lies within the scatter of the total stellar mass density estimates from the observations of \cite{Gonzalez11} (inverted triangles) and \cite{Ilbert13} (open diamonds), although the simulation data is above the estimates of \cite{Muzzin13} (open circles) above redshift 2.  
Between redshifts 2 and 0.1 the simulation data lies within the error bars from different observational estimates, although it is on the lower side of all observed values below redshift 0.9.
At redshift 0.1, where \smd\ can be determined most accurately from observations, the simulation falls below the observations by a small amount, less than 0.1~dex, or 20 per cent. We will return to the source of this deficit in stellar mass at low redshift when studying the shape of the GSMF.

Returning to the agreement between redshifts 2 and 4, above redshift 2 the stellar mass density estimated from observations requires extrapolation below the mass completeness limit of the survey, as discussed.  
To compare the simulation with the stellar mass density that is observed, without extrapolation, the red and green lines in the top panel show \smd\ from the simulation after applying the mass completeness limits of \cite{Ilbert13} and \cite{Muzzin13}, respectively.  
The mass completeness limits applied are listed in Table \ref{table:obsmass}.
The red and green lines should be compared to the filled red diamonds and filled green circles, respectively, showing \smd\ from the observed galaxies without extrapolating below the mass completeness limit.  
Note that 30 pkpc apertures are still applied to the simulated galaxies for this comparison.
When comparing with \cite{Ilbert13}, we find agreement at the level of the observational error bars from redshifts 0.2 to 4.
However, \cite{Muzzin13} find more stellar mass than the simulation after applying the mass completeness limits between redshifts 1.5 and 4.  
This can be understood by noting that the estimated mass completeness limit of \cite{Muzzin13} is higher than that of \cite{Ilbert13} (although both groups use the same survey data), resulting in only the most massive objects being detected at a given redshift. 
These objects are not sufficiently massive in the simulation when compared with the inferred GSMF from observations (without accounting for random or systematic mass errors), as will be shown next.

\subsection{The evolution of the galaxy stellar mass function}
\label{sec:gsmf}

\begin{figure*}
  \centering
  \includegraphics[width=1.0\textwidth]{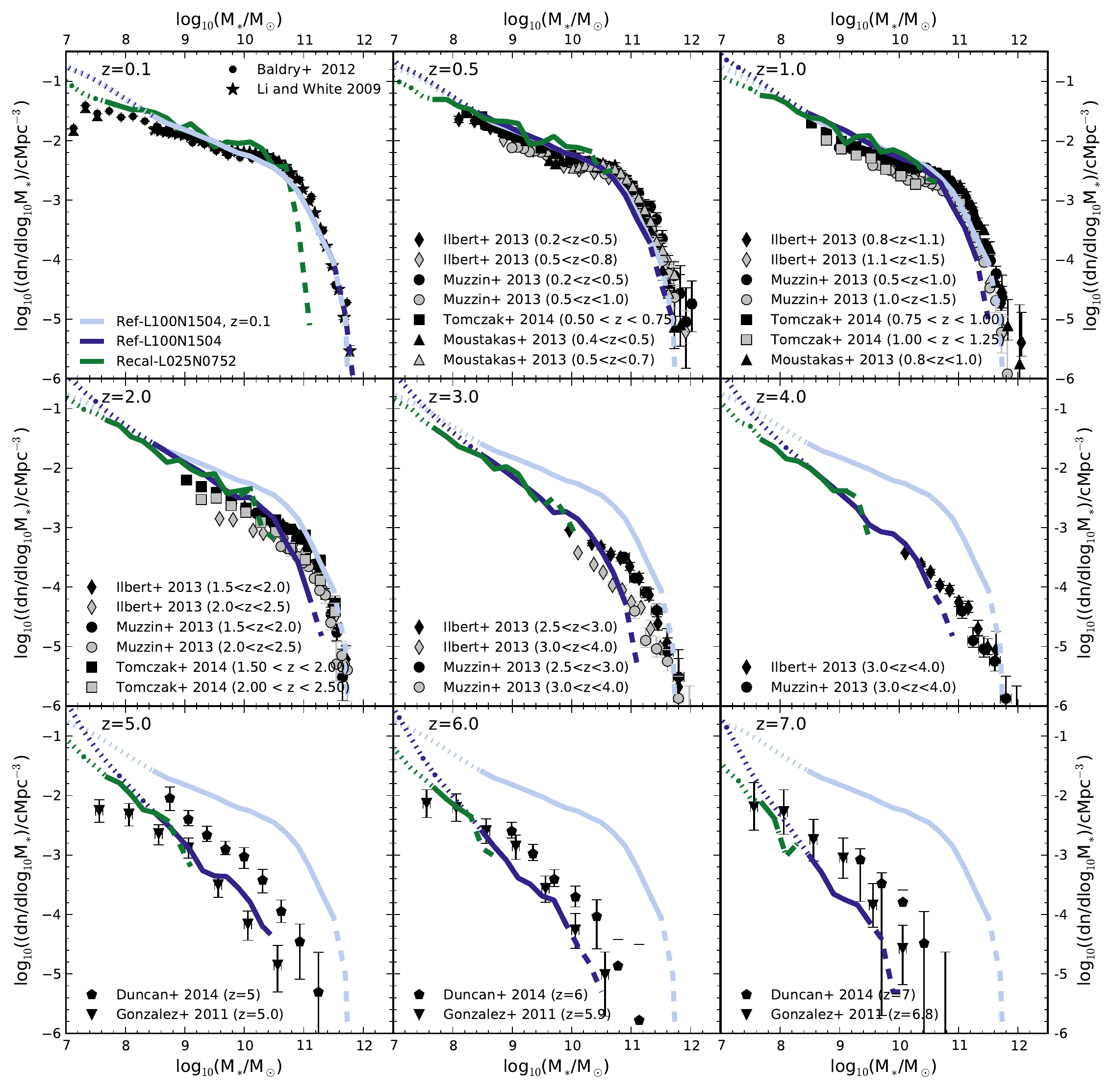}
  \caption{
	The galaxy stellar mass function at the redshifts shown in the upper left of each panel for simulation \bigbox\ and \highres, in blue and green respectively.  
	When the stellar mass falls below the mass of 100 baryonic particles curves are dotted, when there are fewer than 10 galaxies in a stellar mass bin curves are dashed.
	The redshift $0.1$ GSMF is reproduced in each panel as a light blue curve, to highlight the evolution.
	Comparing \bigbox\ to \highres, the simulations show good convergence over the redshift range shown, where there are more than 10 galaxies per bin.
	The data points show observations as indicated in the legends.
	Where necessary, observational data have been converted to a Chabrier IMF and Planck cosmology.
	The black points represent the observational redshift bin
	below the simulation redshift, while the grey curves are from the redshift bin above the simulation snapshot. 
	Within the expected mass errors we find good agreement with observations of the GSMF from redshift
	$0$ to $7$.  
	Between redshifts two and four the model tends to underestimate the masses of the brightest galaxies by around 0.2 dex,
	but these are very sensitive to the stellar mass errors in the observations, see text for discussion.
	}
  \label{fig:smfevo}
\end{figure*}

The evolution of the stellar mass density of the Universe provides a good overview of the growth of stellar mass in the simulation. However, it does not test whether stars form in galaxies of the right mass.
We now carry out a full comparison of the GSMFs in the simulation with those inferred from observations at different epochs.  

The shape of the GSMF is often described by a \cite{Schechter} function,
\begin{equation}
\Phi(M) dM =  \Phi^* \left( \frac{M}{M_{\rm C}} \right) ^{\alpha} e^{-\frac{M}{M_{\rm C}}} dM,
\label{eq:Sch}
\end{equation}
where \Msch\ is the characteristic mass or ``knee'', \Phis\ is the normalisation and \alphas\ is the power-law slope for $M \ll$ \Msch.
We will refer to the slope and knee throughout this comparison.
In Appendix \ref{ap:Sch} we fit the simulation GSMFs with Schechter functions to provide a simple way of characterising the simulated GSMFs.  

In Figure \ref{fig:smfevo} we compare the GSMF to the same observational data sets that were presented in Figure \ref{fig:smdensity} in terms of the total stellar mass density.
The GSMFs from these different observations are consistent with each other within their estimated error bars up to redshift two.
Between redshifts 0 and 1 there is little evolution seen in the observational data, all show a reasonably flat low-mass slope and a normalisation that varies by less than 0.2~dex at $10^{10}$\msun\ over this redshift range. From redshift 1 to 2 there is a steepening of the slope at galaxy masses below $10^{10}$\msun\ and a drop in normalisation of $\sim 0.4$~dex. The drop in normalisation appears to continue above redshift two, although the observations do not probe below $10^{10}$\msun\ at redshifts two to four.

Observational data at redshifts 5, 6 and 7 from \cite{Gonzalez11} and \cite{Duncan14}, based on rest-frame UV observations, are shown in the bottom three panels of Figure \ref{fig:smfevo}.
There is no clear break in the GSMF at these high redshifts, so it is not clear that the distribution is described by a Schechter function in either data set. 
Both data sets show similar slopes above $10^8$\msun.
At low masses, below $10^8$\msun, the data set of \cite{Gonzalez11} shows a flattening in the slope at all redshifts shown.  
These low masses are not probed by \cite{Duncan14}.
At redshift 5 the data sets differ in amplitude by up to 0.8 dex. This offset reduces to $\sim 0.2$~dex by redshift 7. A comparison of these data sets provides an impression of the systematic errors in determining the GSMF from observations at redshifts greater than 5.

We compare these observations to the evolution of the GSMFs predicted by \bigbox\ between redshift $0.1$ and $7$, spanning $13$ Gyr.  The GSMF for \bigbox\ is shown as a blue curve in Figure \ref{fig:smfevo}, and to guide the eye, we repeat the redshift 0.1 GSMF in all panels in light blue. To facilitate a direct comparison with observational data, the GSMF from \bigbox\ is convolved with an estimate of the likely uncertainty in 
observed stellar masses. Random errors in observed masses will skew the shape of the stellar mass function
because more low-mass galaxies are scattered to higher masses than vice versa. We use the uncertainty quoted by \cite{Behroozi13}, $\sigma(z) = \sigma_0 + \sigma_zz$ dex, where $\sigma_0 = 0.07$ and $\sigma_z = 0.04$.
This gives a fractional error in the galaxy stellar mass of 18\% at redshift 0.1 and 40\% at redshift 2.
Note that this error does not account for any systematic uncertainties that arise when inferring the stellar mass from observations, which could range from 0.1 to 0.6~dex depending on redshift (see Section \ref{sec:errors}).

Recall that the observed GSMF at redshift 0.1 was used to calibrate the free parameters of the simulation.
At this redshift, the simulation reproduces the reasonably flat slope of the observed GSMF below $10^{10.5}$\msun, with an exponential turnover at higher masses, between $10^{10.5}$\msun\ and $10^{11}$\msun.  
Overall, we find agreement within 0.2 dex over the mass range from $2 \times 10^8$\msun\ to over $10^{11}$\msun\ and a very similar shape for the simulated and observed GSMF.
In our implementation, the interplay between the subgrid stellar and AGN feedback models at the knee of the GSMF, at galaxy masses of around $10^{10.5}$\msun, results in a slight underabundance of galaxies relative to observations.  
As the stellar mass contained in this mass range dominates the stellar mass density of the Universe, this small offset accounts for the shortfall of stellar mass at the $20\%$ level seen at redshift zero in \smd\ in Figure \ref{fig:smdensity} (blue curve).  

In the simulation, there is almost no evolution in the GSMF from redshift zero to one, apart from a small decrease of 0.2 dex in galaxy masses at the very high-mass end.  
This can be seen by comparing the blue and light blue lines in the top panels, where the light blue line repeats the redshift 0.1 GSMF.  
A similar minimal evolution was reported based on the observational data of \cite{Moustakas13}(triangles) from redshift 0 to 1, and is also seen in the other data sets shown. 

From redshift one to two the simulation predicts strong evolution in the GSMF, in terms of its normalisation, low-mass slope and the location of the break.  
Between these redshifts, spanning just 2.6 Gyr in time, the stellar mass density almost doubles, from 0.75 to 1.4 $\times 10^8$\msun cMpc$^{-3}$, and the GSMF evolves significantly.
From redshift two to four the normalisation continues to drop and the mass corresponding to the break in the GSMF continues to decrease.

Although the trend of a decrease in normalisation of the GSMF between redshift one and two is qualitatively consistent with what is seen in the observations, the normalisation at redshift two at $10^{9.5}$\msun\ is too high in the simulation by around 0.2 dex. 
There is also a suggestion that the normalisation of the GSMF in the simulation is too high at redshift three, although observations do not probe below $10^{10}$\msun\ at this redshift.
It is therefore difficult to draw a strong conclusion from a comparison above redshift 2 without extrapolating the observational data.
At redshift two there is also an offset at the massive end of the GSMF.  
The exponential break occurs at a mass that is around 0.2 dex lower than observed.  
However, the number of objects per bin in the simulation at redshift two above $10^{11}$\msun\ falls below 10 providing a poor statistical sample of the massive galaxy population.
Increasing the box size may systematically boost the abundance of rare objects, such as that of galaxies above $10^{11}$\msun\ at redshift two and above.
The break is also particularly sensitive to any errors in the stellar mass estimates, a point we will return to below.

Comparing the simulated GSMF to observations at redshifts 5, 6 and 7, we find a similar shape to the observational data.
The simulation has a similar trend with mass to \cite{Gonzalez11}, however it is offset in stellar mass from \cite{Duncan14}.
No break in the GSMF is visible, neither in the simulation nor in the observations, at these high redshifts over the mass ranges considered here. 
Hence, for redshifts above 5 a Schechter fit may not be an appropriate description of the data.

\subsubsection{Galaxy stellar mass errors}
\label{sec:errors}

\begin{figure}
  \centering
  \includegraphics[width=0.5\textwidth]{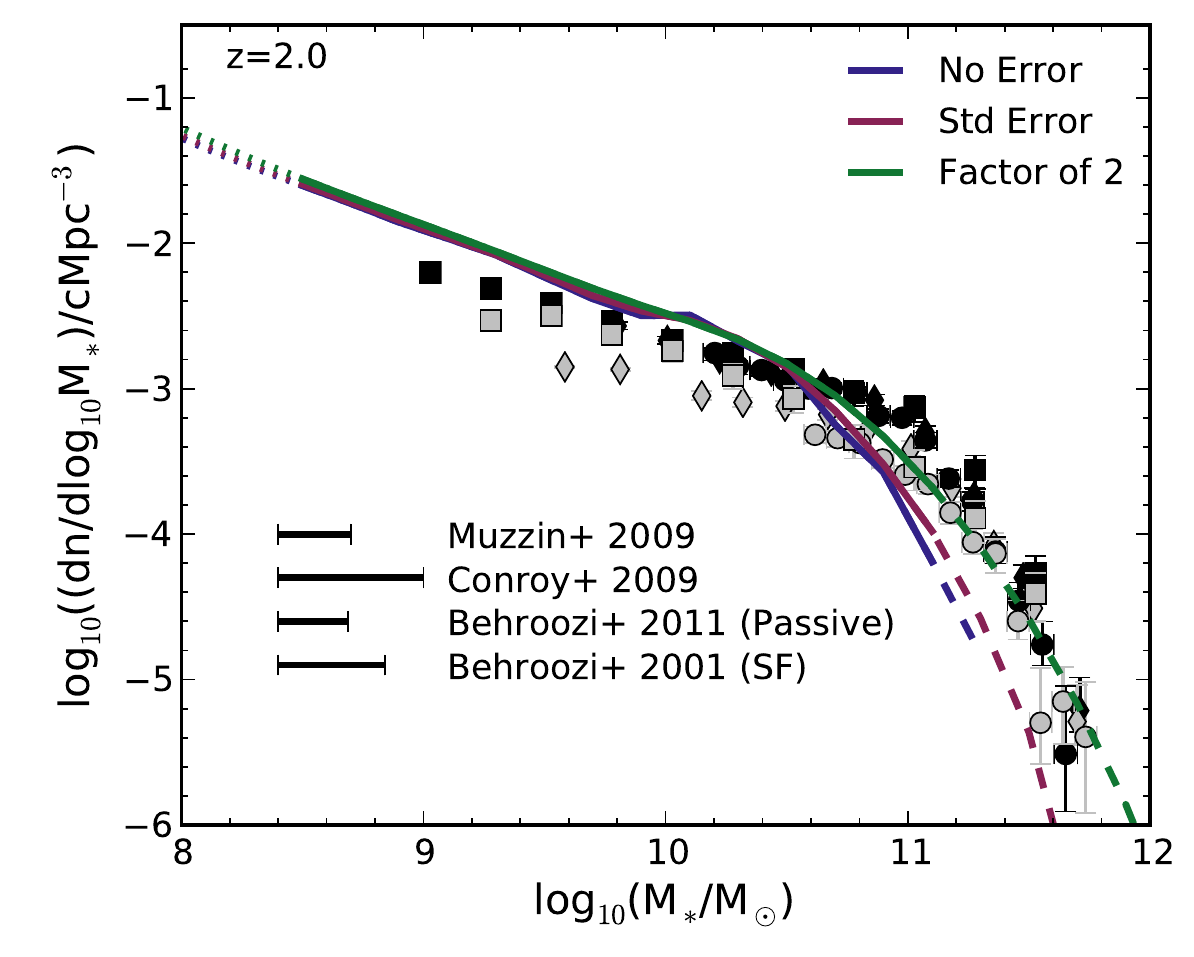}
  \caption{
 The simulated GSMF at redshift two from \Eagle\ without random mass errors (red), convolved with the stellar mass error of \citet{Behroozi13}, used in Figure \ref{fig:smfevo}, (blue) and with random errors of a factor two (green).  
 The random errors have a significant effect on the shape of the massive end of the GSMF, transforming the simulation from mildly discrepant with the observational data to being in excellent agreement with data.  
 The Gaussian convolution with a stellar mass error is motivated by the random errors associated with the Malmquist bias.
 The horizontal black lines in the lower left of the figure indicate the estimated magnitudes of systematic errors in stellar masses according to \citet{Muzzin09}, \citet{Conroy09} and \citet{Behroozi13} at redshift two.
 Systematic errors are expected to maintain the shape of the GSMF but would shift it horizontally.
 Within the estimated level of uncertainty in observations, the simulation shows agreement with observations of the GSMF, including the location of the break, although the low-mass slope may be slightly too steep.
}
  \label{fig:masserrors}
\end{figure}

When comparing the simulation to observations, it is important to consider the role of stellar mass errors, both random and systematic.
We begin by considering the random errors.
In Figure \ref{fig:masserrors} the GSMF from \bigbox\ is plotted at redshift two assuming no stellar mass error (red), a random mass error of $0.07+0.04z$ \citep{Behroozi13} as in Figure \ref{fig:smfevo} (blue), resulting in an error of 40\% in galaxy stellar mass at redshift two,  and a mass error of a factor of two (green), i.e. 100\%. 
Where the GSMF is reasonably flat, i.e. at masses below $10^{10.5}$\msun, the impact of random uncertainty is minimal.
However, above this mass the shape of the GSMF depends strongly on the random stellar mass errors in the observations, because more low-mass galaxies are scattered to high masses than vice versa. 
If we increase the random errors, the exponential break becomes less sharp and the simulation agrees better with the observations.

There are also systematic errors to consider in the determination of stellar masses from observed flux or spectra.
Fitting the spectral energy distribution (SED) of a galaxy is sensitive to the choice of stellar population synthesis (SPS) model, e.g. due to the uncertainty in how to treat TP-AGB stars, the choice of dust model and the modelling of the star formation histories \citep[e.g.][]{Mitchell13}.
Systematic variations in the stellar IMF would result in additional uncertainties, which are not considered here.
The systematic uncertainties from SED modelling increase with redshift.
At redshift zero \cite{Taylor11} quote $\sim 0.1$ dex ($1\sigma$) errors for GAMA data.
At redshift two the estimated systematic error on stellar masses ranges from 0.3 dex \citep{Muzzin09} to 0.6 dex \citep{Conroy09}, based on uncertainties in SPS models, dust and metallicities.
Figure \ref{fig:masserrors} gives an impression of the size of these systematic errors by plotting values from \cite{Muzzin09}, \cite{Conroy09} and \cite{Behroozi13} in the bottom left corner.
The \cite{Behroozi13} estimate is divided into star forming and passive galaxies due to the reduced sensitivity of passive galaxies to the assumed form of the star formation history.
The systematic stellar mass errors are expected to shift the GSMF along the stellar mass axis.
Considering the extent of the systematic uncertainties, we find the GSMF from \Eagle\ to be consistent with the observational data, although the low-mass slope may be slightly too steep.
The observed evolutionary trends in the normalisation and break are reproduced by the simulation, suggesting that the simulation is reasonably representative of the observed Universe.

\subsubsection{Numerical convergence}
\label{sec:weakcon}

Having found reasonable agreement between the evolution in the \bigbox\ simulation and the observations, it is important to ask if the results are sensitive to numerical resolution.
We consider only weak convergence tests here, {\it i.e.} we only examine the ability of the simulation to reproduce the observed evolution after recalibrating the high-resolution simulation to the same conditions (namely the redshift 0.1 GSMF) as used for the standard resolution simulation.
In Figure \ref{fig:smfevo} the high-resolution model, \highres, is shown in green.

The 25 cMpc box is too small to sample the break in the GSMF accurately.
To avoid box size issues, we do not consider the GSMF when there are fewer than 10 galaxies per bin, i.e. where the green curve is dashed.   The 25 cMpc box also shows more fluctuations, due to poorer sampling of the large-scale modes in a smaller computational volume.  At masses below $10^8$\msun, when there are fewer than 100 star particles per galaxies in the \bigbox\ simulation (blue dotted curve), the slope of the high-resolution simulation is flatter than that of \bigbox.
Where the solid part of the blue and green curves overlap, there is excellent agreement, to better than 0.1~dex, between both resolutions across all redshifts.  Overall, this amounts to good (weak) numerical convergence in the simulation across all redshifts that can be probed, given the limitations imposed on the test due to the small volume of the high-resolution run.

\vspace{10mm}

In summary, we have found the stellar mass density in the simulation to be close to the values estimated from observations, with a maximum offset of $\sim 20\%$ due to the slight undershooting of the \eagle\ GSMF around the knee of the mass function. The observed evolutionary trends, in terms of changes in the shape and normalisation of the GSMF between redshift 0.1 to 7 are reproduced, although the evolution in the normalisation is not sufficiently strong in the simulation from redshift 1 to 2, with an offset in normalisation at redshift 2 of $\sim 0.2$~dex. The break in the GSMF occurs at too low a mass in the simulation compared to the observations at redshifts 2 to 4. However, the box size limits the number of objects produced in the simulation and we have shown that stellar mass errors play a significant role in defining the observed break of the GSMF. As a result of these uncertainties affecting the comparison, 
the remaining differences between the simulation and observations do not suggest significant discrepancies in the model.

\section{Evolution of star formation rates}
\label{sec:sfr}

\begin{figure}
  \centering
  \includegraphics[width=0.5\textwidth]{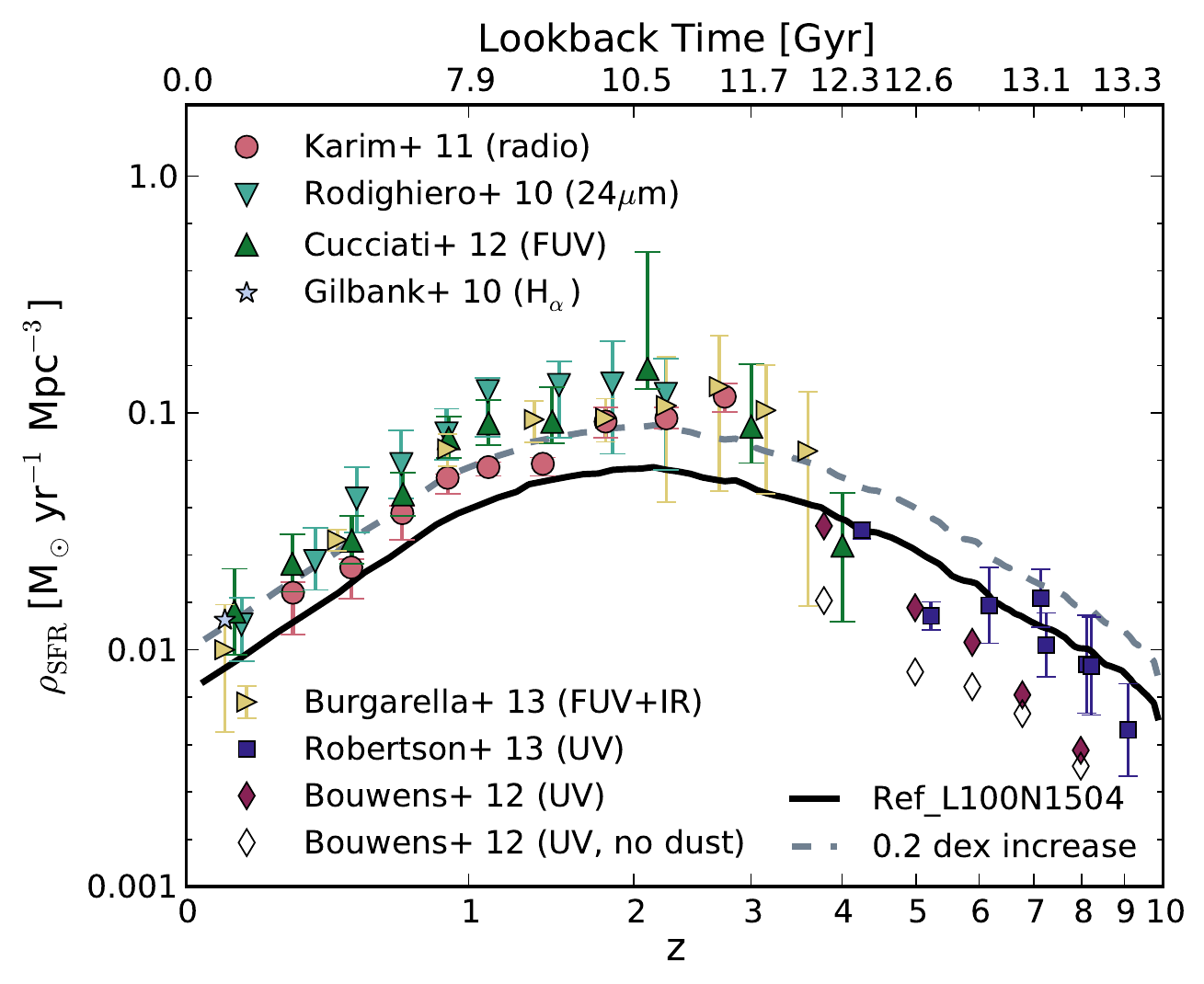}
  \caption{ 
	Evolution of the cosmic star formation rate density. The \Eagle\ simulation \bigbox\ is plotted as a solid black curve, observational data are plotted as symbols. Open symbols from \protect \cite{Bouwens12} exclude a dust correction to the SFRs, giving an impression of the uncertainty in the measurement.
The simulation tracks the evolution of the observed \sfrd\ very well, albeit with an almost
constant 0.2~dex offset (grey dashed line) below redshift $z\sim 3$.}
  \label{fig:sfrh}
\end{figure}

\subsection{The cosmic star formation rate density}
\label{sec:sfrh}
The star formation rate density (\sfrd) as a function of redshift is plotted for simulation \bigbox\ in Figure \ref{fig:sfrh}. For comparison, observations from \cite{Gilbank10} $\lbrack$H$\alpha\rbrack$, \cite{Rodighiero10} $\lbrack$24$\mu$m$\rbrack$, \cite{Karim11} $\lbrack$Radio$\rbrack$, \cite{Cucciati12} $\lbrack$FUV$\rbrack$, \cite{Bouwens12} $\lbrack$UV$\rbrack$ , \cite{Robertson13} $\lbrack$UV$\rbrack$ and \cite{Burgarella13} $\lbrack$FUV + FIR$\rbrack$ are shown as well. This compilation of data covers a number of SFR tracers, providing an overview of \sfrd\ estimates from the literature, as well as an indication of the range of scatter and uncertainty arising from different methods of inferring \sfrd.
There is a spread in the measured \sfrd\ of around $0.2$~dex at redshifts less than two, while the estimated \sfrd\ include error bars of about $\pm 0.15$ dex, with larger error bars above redshift two. 

At high redshift the simulated \sfrd\ (solid black curve) increases with time, peaks around redshift two, followed by a decline of almost an order of magnitude to redshift zero.  The simulation reproduces the shape of the observed \sfrd\ as a function of time very well, but falls below the measurements by an almost constant and small offset of 0.2~dex at $z \le 3$. (The grey dashed line in Fig. \ref{fig:sfrh} shows \sfrd\ increased by 0.2 dex.)
While the simulation agrees reasonably well with the observational data at redshifts above 3, we caution that these measurements are reasonably uncertain.
For example, the difference between open and filled symbols for \cite{Bouwens12} data shows the estimated dust correction that is applied to the observations.

\begin{figure*}
  \centering
  \includegraphics[width=1.0\textwidth]{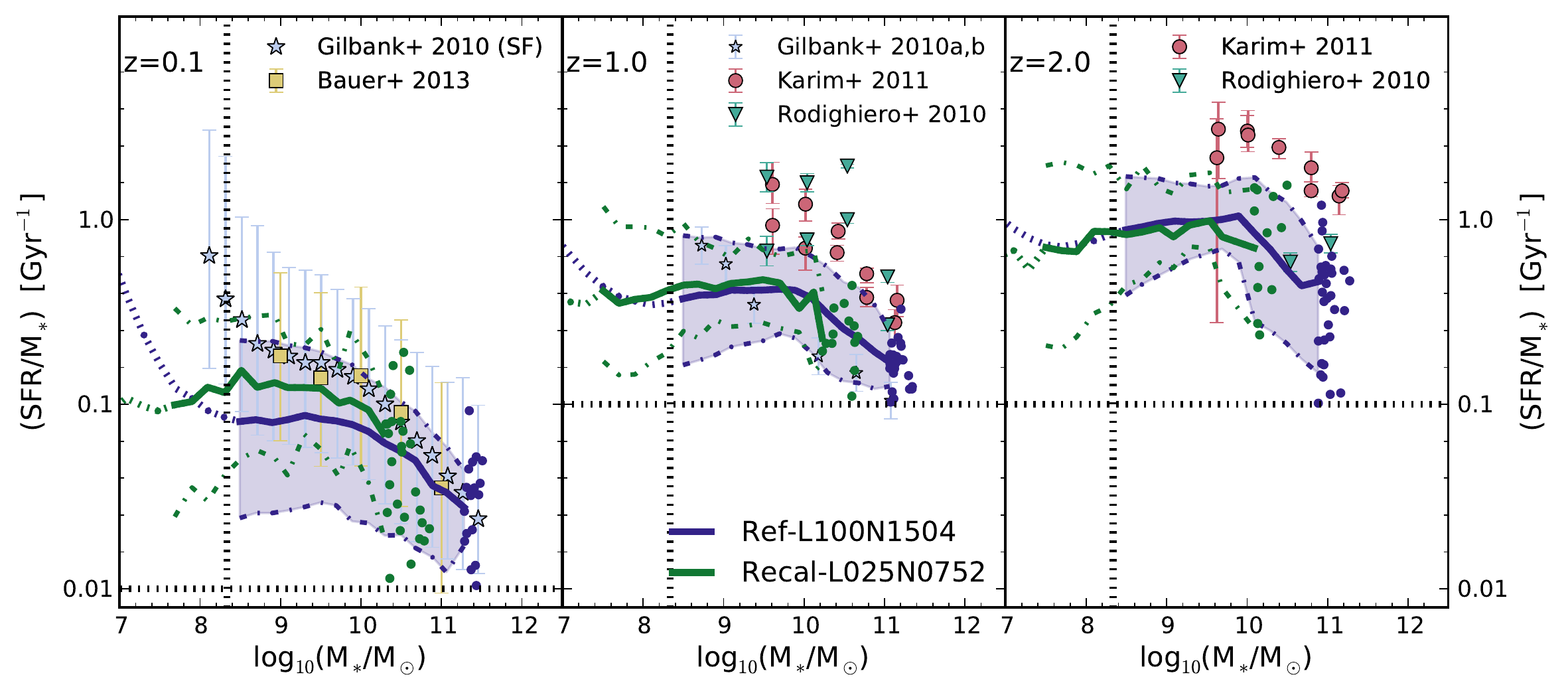}
  \caption{
	The specific star formation rate (SSFR), $\dot{M_*}$/$M_*$, as a function of galaxy stellar mass for \bigbox\ and \highres\ from left to right at redshifts 0.1, 1 and 2.
	The solid curves show the median relation for star forming galaxies, defined as those with a SSFR above the limit specified by the horizontal dotted line in each panel.
	The shaded region (dot dashed curves) encloses the 10$^{\rm th}$ to 90$^{\rm th}$ percentiles for \bigbox\ (\highres).
	Where there are fewer than 10 galaxies per bin, individual data points are shown.
	Lines are dotted when the stellar mass falls below that corresponding to 100 star-forming particles for the median SSFR and the mass of 100 baryonic particles, to indicate that resolution effects may be important.
	At redshift $0.1$ the observational of \citet{Gilbank10} and \citet{Bauer13} are shown as light blue stars and yellow squares, respectively.
	Error bars enclose the $10^{\rm th}$ to $90^{\rm th}$ percentiles.  
	At higher redshift, data from \citet{Gilbank10b}, \citet{Karim11} and \citet{Rodighiero10} are shown as light blue stars, pink circles and turquoise inverted triangles respectively.
	The observed flat slope with stellar mass and the increase in normalisation with redshift are reproduced by the simulations, but the simulation is lower in normalisation by 0.2 to 0.4 dex, depending on redshift and the observational data set.
	}
	\label{fig:ssfrevo}
\end{figure*}

\begin{figure*}
  \centering
  \includegraphics[width=1.0\textwidth]{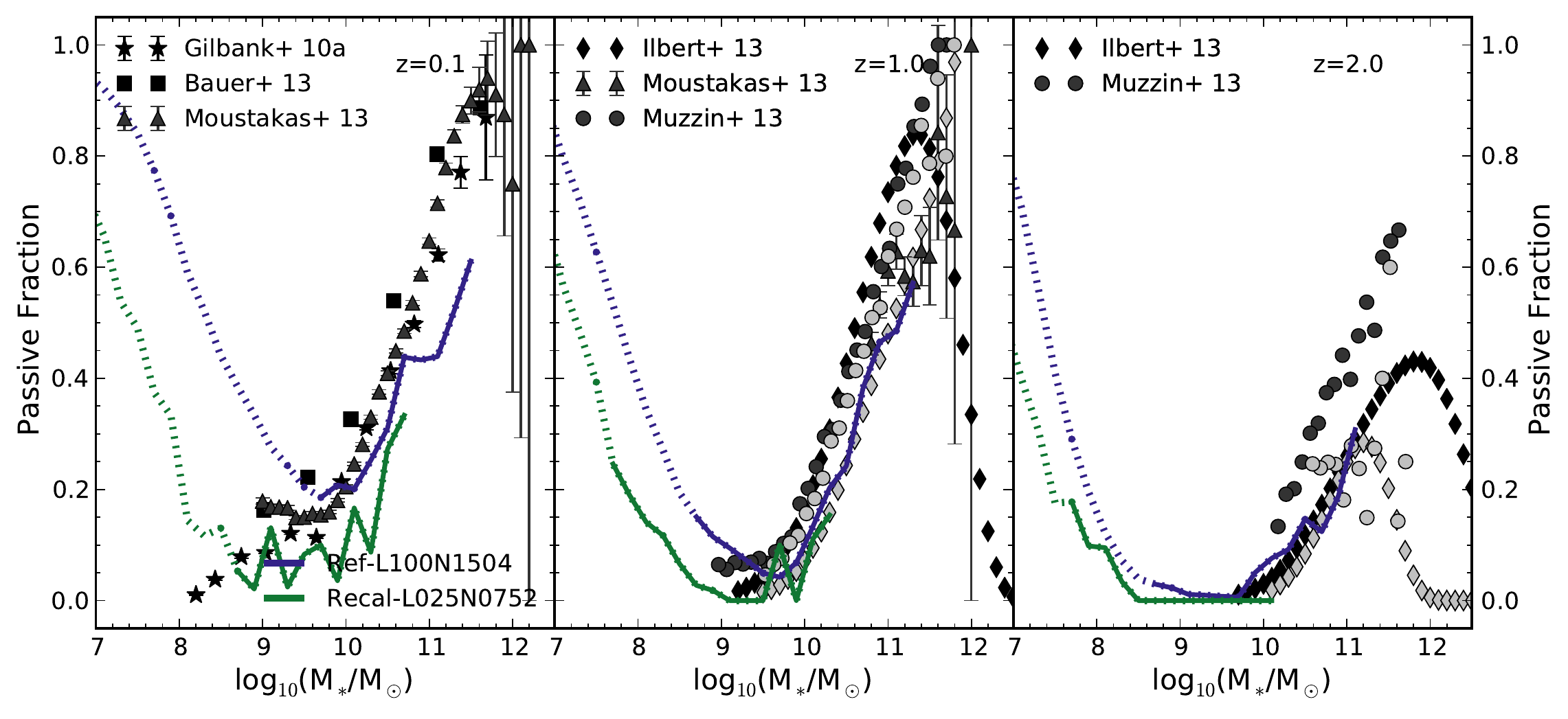}
  \caption{
	The passive fraction as a function of galaxy stellar mass for \bigbox\ and \highres\, in blue and green, respectively, 
	where galaxies with a SSFR below the horizontal dotted lines in Figure \ref{fig:ssfrevo} are defined as passive.  
	Lines are dotted when the stellar mass falls below that corresponding to 30 star-forming particles for the SSFR limit.
	Data points show observations as indicated in the legends.
	The black points represent the observational redshift bin
	below the simulation redshift, while the grey curves are from the redshift bin above the simulation snapshot. 
	Above $10^9$\msun, the simulated passive fractions show similar normalisation and slope with stellar mass to observations at all redshifts, with a small deficit of passive galaxies of around 15\% in the mass range $10^{10.5}$ to $10^{11.5}$\msun.
	The upturn at low masses, below $10^9$\msun\ is a numerical artefact.
	}
	\label{fig:passive}
\end{figure*}

\begin{figure*}
  \centering
  \includegraphics[width=1.0\textwidth]{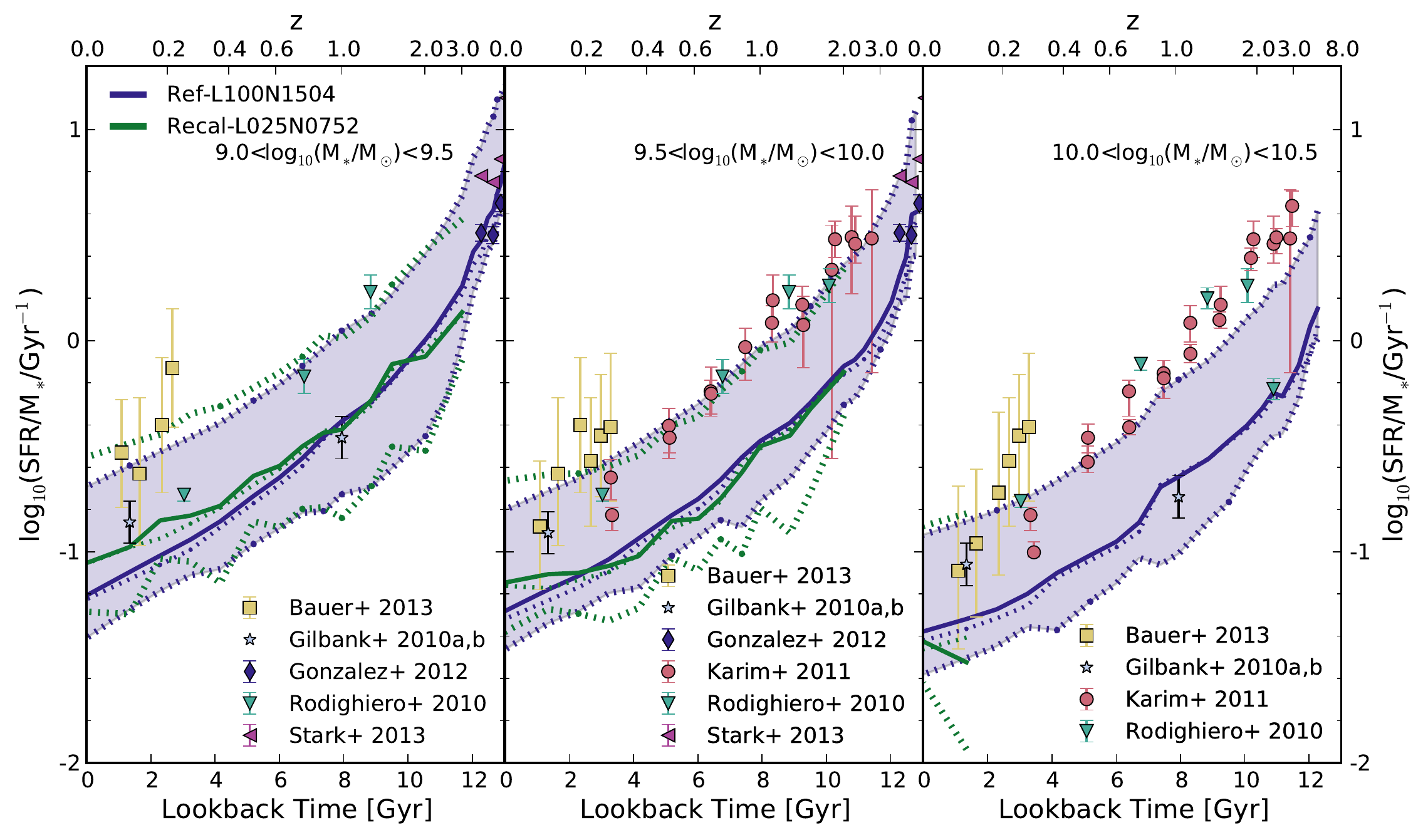}
  \caption{Evolution of the specific star formation rate (SSFR) as a function of lookback time for
    stellar mass bins $10^{9.0} < $M$_* < 10^{9.5}$\msun\ (left), $10^{9.5} < $M$_* < 10^{10.0}$\msun\ (middle) 
	and $10^{10.0} < $M$_* < 10^{10.5}$\msun\ (right) for \bigbox\ and \highres, in blue and green, respectively. 
    Solid curves show the median SSFR from the simulation for star forming galaxies, the shaded region (dotted curves) enclose the $10^{\rm th}$ and $90^{\rm th}$ percentile values for \bigbox\ (\highres).
	Medians are only shown when there are more than 10 galaxies per bin.
	Observational data from \citet{Gilbank11}, \citet{Bauer13}, \citet{Karim11}, \citet{Rodighiero10}, \citet{Gonzalez12} and \citet{Stark13} are shown.
	The simulation shows good agreement with the observed shape of the SSFR evolution, but there is an offset in normalisation of 0.2 to 0.4 dex, as seen in Figure \ref{fig:ssfrevo}.
	}
  \label{fig:ssfrz}
\end{figure*}

\subsection{Specific star formation rates}
\label{sec:ssfr}
Observationally, a well defined star forming sequence as a function of stellar mass has been found in the local Universe, which appears to hold up to a redshift of 3 \citep[e.g][]{Noeske07, Karim11}.
It is described by a relation of the form
\begin{eqnarray}
\frac{\dot{\rm M_*}}{\rm M_*} = \beta \left(\frac{\rm M_*}{10^{10}\rm M_\odot}\right)^\gamma\,,
\end{eqnarray}
where \gammas\ is the logarithmic slope, \betas\ is the normalisation and $\dot{\rm M_*}/\rm M_*$ is the specific star formation rate (SSFR).
Observations indicate that \gammas\ is negative but close to zero, and it is often assumed to be constant with stellar mass.

Figure \ref{fig:ssfrevo} shows the SSFR for star forming galaxies as a function of galaxy stellar mass at redshifts 0.1, 1 and 2.  
The observational data sets for the SSFRs we compare to at redshift 0.1 are from \cite{Gilbank10} (stars) and \cite{Bauer13} (squares).
These data sets show similar values for the normalisation and slope and a similar scatter above $10^9$\msun.
Below $10^9$\msun\ only \cite{Gilbank10} data is available.  This data shows an increase in the SSFR with decreasing stellar mass below $10^{8.5}$\msun.
\cite{Rodighiero10} (inverted triangles), \cite{Karim11} (circles) and \cite{Gilbank10b} (stars) are shown at higher redshifts.
Comparing these data sets, \cite{Rodighiero10} and \cite{Karim11} have similar slopes and normalisation at redshifts one and two.  
However, the \cite{Gilbank10b} data is substantially (0.8 dex) lower in normalisation over the mass ranges where it overlaps with \cite{Rodighiero10} and \cite{Karim11}. 
The ROLES data used by \cite{Gilbank10b} probes faint galaxies down to masses below $10^9$\msun, but this deep survey covers only a small area of sky.
The resulting small number statistics of massive galaxies may be driving this offset in SSFR from the other observational data sets.

The median SSFRs for star forming galaxies from \bigbox\ and \highres\ are shown as blue and green curves, respectively.
The horizontal dotted lines correspond to the SSFR cut ($\sim 1$ dex below the observational data) used to separate star forming from passive galaxies.

At redshift 0.1 the SSFR in the simulations is reasonably independent of stellar mass (where well resolved) up to masses of $10^{10}$\msun.
Above this mass the SSFR decreases slowly with stellar mass.
The simulations show a scatter of around 0.6 dex across the stellar mass range resolved by \bigbox.
The normalisation of the \highres\ simulation lies 0.2 dex above that of \bigbox, as was already shown in \Schaye.
At low masses, when there are fewer than 100 star-forming particles per galaxy, there is an increase in SSFR with stellar mass in \bigbox.
However, by comparing with \highres\ we see that this is resolution driven.

The trend with stellar mass above $10^9$\msun\ is similar in the simulations and the observations.
However, there is an offset in the normalisation from observations, where \highres\ and \bigbox\ are low by $\sim 0.1$ and 0.3 dex respectively.
The increase in SSFR at a stellar mass of $10^{8.5}$\msun\ reported by \cite{Gilbank10} is not seen in the \highres\ simulation, which has sufficient numerical resolution to compare to observations at these low masses.
This could indicate that stellar feedback is too strong in low-mass galaxies, or perhaps that the observational data is not volume complete due to the difficulty in detecting low-mass galaxies with low star formation rates owing to their low surface brightness (see \Schaye\ for more discussion of the redshift 0.1 properties).

At higher redshifts the simulation SSFRs increase in normalisation, maintaining a flat slope below $10^{10}$\msun, with a shallow negative slope above this stellar mass.
At redshifts between one and two the \highres\ and \bigbox\ SSFRs lie within 0.1 dex of each other across the stellar mass ranges for which both are resolved.
The increase in normalisation seen in the simulations reproduces the observed trend, although the offset in normalisation increases to up to 0.5 dex when comparing to the data sets of \cite{Rodighiero10} and \cite{Karim11}. 
Relative to the \cite{Gilbank10b} data at redshift one, the median SSFR from the simulation agrees to within around 0.2 dex.
Comparing the slope of the SSFR-M$_*$ relation of \cite{Gilbank10b} to the simulations, the simulation is flatter below $10^{10}$\msun, but is in agreement with the slopes of \cite{Karim11} and \cite{Rodighiero10}.

Observationally the galaxy population exhibits a bimodal colour distribution, which may imply a bimodality in the SSFR. 
To study this bimodality in the simulation, we show in Figure \ref{fig:passive} the passive fraction of galaxies as a function of mass at redshifts 0.1, 1 and 2.
At higher redshifts the simulation volume does not provide sufficiently massive galaxies to overlap with those detectable in observations.
In the simulation we define passive galaxies by a cut in SSFR that is an order of magnitude below the median observed SSFR (dotted horizontal line in Figure \ref{fig:ssfrevo}).
Varying this limit, while keeping it below the main star forming sequence has negligible impact on the recovered median SSFR, although it can increase or decrease the passive fractions by around 10\%.

For comparison, passive fractions from \cite{Gilbank10}, \cite{Bauer13} and \cite{Moustakas13} are shown at redshift $0.1$ and from \cite{Moustakas13}, \cite{Muzzin13} and \cite{Ilbert13} at higher redshifts. 
For most observational data sets shown, the passive fraction is determined based on a colour or SSFR cut as applied in the published data sets.
\cite{Gilbank10} provide tabulated stellar masses and SFRs for each galaxy and we therefore apply the same SSFR cut as we use for the simulation data.  
At redshift 0.1 the dependence of passive fraction on stellar mass is similar for all observational data sets.  At redshift one, each observational data set shows the same trend, but there is a difference of up to 0.15 in the passive fraction for M$_* \lesssim 10^{11}$\msun\ for different data sets, and a larger difference above this mass.
At redshift two agreement between data sets is poor.

The passive fraction from \bigbox\ and \highres\ are shown in blue and green, respectively.
As a resolution guide, where the stellar mass is less than the maximum of 100 baryonic particles and 30 gas particles for the mass that corresponds to the SSFR cut, lines are dotted.
As the SSFR cut evolves with redshift, this resolution guide evolves with redshift.
The guide was chosen based on a comparison of the passive fractions for central galaxies in \bigbox\ and \highres\ (not shown).
Both feedback and environment can quench star formation in galaxies.  As different environments are probed in simulations of different box size, the passive fractions are expected to differ between \bigbox\ and \highres, not only because of the resolution but also due to the box size. 
To overcome this, a comparison is carried out for central galaxies in the two simulations, which probe similar environments.
This yields a difference in the passive fractions when a galaxy's stellar mass is resolved by a minimum of 100 particles and the SSFR for the passive threshold is resolved by a minimum of 30 gas particles.
 
Over the resolved mass range, the passive fraction at redshift 0.1 follows a similar trend to the observational data, although there are too few passive galaxies between $10^{10.5}$ and $10^{11.5}$\msun\ by around 15\%.  
In the simulations, passive fractions are lower at redshift 1 than at redshift 0.1
This is consistent with what is seen in observational studies, although, there are again fewer passive galaxies in the range of $10^{10.5}$ to $10^{11.5}$\msun\ than observed.
At redshift two there is a further drop in the passive fraction of galaxies, both in the simulation and the observations.  
Summarising, the passive fractions show the same trend as observations when galaxy masses and SFRs are resolved, although there are too few passive galaxies by $\sim 15\%$ in the stellar mass range $10^{10.5}$ to $10^{11.5}$\msun.

To better study the evolution of the SSFR and to extend the comparison to higher redshifts, we show in Figure \ref{fig:ssfrz} the SSFR as a function of lookback time in three different stellar mass bins, of 0.5 dex centred on $10^{9.25}$, $10^{9.75}$ and $10^{10.25}$\msun.
The median SSFR for star forming galaxies from \bigbox\ and \highres\ are shown in blue and green, respectively. 
In all mass bins the SSFR increases with lookback time.
Comparing the two simulation, above redshift one the SSFRs of the two simulations are converged to within 0.1 dex. At lower redshifts, for stellar masses below $10^{9.5}$\msun\ \highres\ has a slightly higher SSFR, by up to 0.2 dex.
Similar trends are found when considering other mass bins of 0.5 dex between $10^{8.5}$ and $10^{11.5}$\msun.

%Values are only shown when there are more than 10 galaxies per bin. 
%The 10$^{\rm th}$ to 90$^{\rm th}$ percentiles are enclosed by the shaded region (dotted curves) for \bigbox\ (\highres). 
%The median SSFR from \bigbox\ in the first panel is reproduced in subsequent panels as a dashed curve to emphasise trends with stellar mass. 

We compare the simulation data with the observations presented in Figure \ref{fig:ssfrevo}, adding \cite{Gonzalez12} and \cite{Stark13} at redshifts 4 and above.
The observed trend with redshift is reproduced, there is, however, an offset in normalisation of $0.2 - 0.5$ dex at all times, across all mass ranges, as seen in Figure \ref{fig:ssfrevo}.
We found previously that the global star formation rate density was low by $\sim 0.2$ dex across all redshifts relative to the values estimated from observations (Section \ref{sec:sfrh}).
An offset in \sfrd\ does not convert directly into an offset in SSFR, due to the potential increase in stellar mass if SFRs were to increase.
The offset in \sfrd\ thus can not fully account for the offset in SSFR.
If the SFRs were boosted by 0.3 dex across all mass ranges at all redshifts, as required to produce more consistent results relative to the observational data, the agreement for the stellar mass density from Section \ref{sec:smdensity} would be broken.
A possible solution to the low SSFRs is that the star formation in the simulated galaxies is not sufficiently bursty. 
We will return to this possibility in the discussion.

As for the stellar mass, there are also uncertainties in the SFRs inferred from observations.
Differences in the measured star formation rate density from different star formation tracers are of order 0.2 dex (as  in Figure \ref{fig:sfrh}), while \cite{Utomo14} claim that SFRs inferred from UV and IR observations may be overestimated relative to those obtained by simultaneously modelling of stellar and dust emission simultaneously.
A recent study by \cite{Boquien14} also find SFRs to be overestimated, in FUV and U bands.
Attempting to quantify the level of uncertainty in SFRs is difficult owing to the different sensitivity of each star formation tracer.
UV observations require a large correction for the light that is absorbed.
IR observations require information about the peak of the SED to constrain the total infrared luminosity and must assume all star formation is shrouded in dust if information from the UV is unavailable.
Radio (and IR) observations can suffer from contamination by AGN and rely on an empirical calibration between the flux and SFR.
At high redshift, where stacking is often necessary due to decreased ability to detect individual objects, there is a risk that the sample is incomplete, biasing results towards higher star formation rates.
\cite{PG08} quote a factor of two (0.3 dex) in the uncertainty of IR SFRs due to dust, \cite{Muzzin09} find a scatter of a factor of 2.8 (0.45 dex) depending on the bands available for fitting the SED.
The SSFRs from the \eagle\ \bigbox\ model are only consistent with observations if the values inferred from the data are systematically high by about a factor of two.

The systematic offset in SSFRs between models and observations has been noted before. 
\cite{Weinmann12} and \cite{Genel14} reported this issue for hydrodynamical simulations, while recent studies such as \cite{Mitchell14} and \cite{White14} revisited the issue with semi-analytic models. 
\cite{White14} propose two plausible solutions to the issue based on their semi-analytic modelling. 
In the first solution star formation in low-mass galaxies forming at early times is preferentially suppressed, delaying star formation and providing further fuel for stars to form at later times. 
In the simulations presented here, \bigbox\ and \highres, the dependence of the feedback on local gas metallicity and density does indeed result in preferential suppression of low mass galaxies at early times and this does improve the behaviour of the SSFRs relative to \Eagle\ simulations with constant feedback or velocity dispersion dependent feedback \citep[presented in][]{Crain15}.
However, to fully resolve the offset in SSFRs much stronger feedback is required in low-mass, high-redshift galaxies than the feedback that is implemented here. 
(Although the requirement for more efficient feedback may in part be a result of numerical radiative losses.)
The second solution that \cite{White14} appeal to, with a similar solution proposed by \cite{Mitchell14}, is limiting the cold gas available for star formation by reducing the accretion of gas from hot and ejected reservoirs onto halos \citep[see also][]{Bower12}. 
As our simulation follows the gravity and hydrodynamics of the gas, it is not a reasonable solution to apply to the accretion of gas in hydrodynamical simulation.

\vspace{10mm}
In summary, the simulation reproduces the shape of the evolution of \sfrd\ with redshift seen in observations with a 0.2 dex offset.
The bimodality in SSFR, the slope with mass and the shape of the evolution of the SSFRs as a function of time are also reproduced by the simulation.
However, the normalisation is 0.2-0.5 dex too low at all redshifts and across all masses. 
This offset cannot be resolved by a simple systematic shift in SFRs in the simulation due to the implications such a shift would have for \smd. 
However the level of uncertainty in the data is such that the level of inconsistency in the \eagle\ specific star formation rates may be smaller than suggested by current observations.

\section{Discussion}
\label{sec:discussion}

We have presented the evolution of the stellar masses and star formation rates in two of the \Eagle\ cosmological hydrodynamical simulations.
We have focused on \bigbox, a ($100$ cMpc)$^3$ box with baryonic particle masses of $1.81 \times 10^6$\msun, and \highres, a (25 cMpc)$^3$ box with baryonic particle masses of $2.26 \times 10^5$\msun.
These simulations use advanced SPH techniques and  state-of-the-art subgrid models, including cooling, metal enrichment, energy input from stellar feedback, black hole growth and feedback from AGN. 
The subgrid parameters depend only on local gas properties.
The free parameters of the model have been calibrated to reproduce the observed local Universe GSMF, with consideration given to galaxy sizes \citet{Crain15}.
The resulting model has been shown to reproduce many observations around redshift zero, including the Tully-Fisher relation, specific star formation rates, the mass-metallicity relation, black hole masses and the column density distribution functions of intergalactic CIV and OVI (\Schaye).

In this paper we extend the comparison with observations of galaxy stellar masses and star formation rates from redshift zero to redshift seven.
This comparison with observations enables us to carry out a multi-epoch verification of the \Eagle\ galaxy formation model, where the galaxy properties in this  comparison are predictions of the model, i.e. evolution histories were not considered during the calibration of model parameters.

We began our comparison by finding a better than 20 per cent agreement with the evolution of the stellar mass density across all epochs (Figure \ref{fig:smdensity}).
For the GSMF, good agreement was typically found for the evolution of the normalisation and break when comparing the simulation to observationally inferred data (Figure \ref{fig:smfevo}).
The normalisation remains reasonably constant from redshift 0.1 to 1 and then decreases to redshift 2.
The decrease continues at higher redshifts.
Although this behaviour is qualitatively consistent with observations, at redshift 2 the normalisation below $10^{10.5}$\msun\ is too high by $\sim 0.2$ dex.
Semi-analytical models have also reported normalisations that are too high relative to observations at $z \sim 2$ \citep[e.g.][although see \cite{Henriques15} for a possible solution in semi-analytics]{Weinmann12}.
In the current \Eagle\ implementation of stellar feedback, galaxies with low metallicity and high density, typical in the early universe, experience strong feedback.
The available feedback energy can be up to three times that available from core collapse supernova, which compensates for numerical radiative losses.
A comparison with the normalisation of the observed GSMF at redshift 2 suggests that even more efficient stellar feedback is required in low mass objects at redshifts above two.
More efficient feedback at high redshifts could provide surplus gas at later times, through recycling, helping to boost the SSFRs ($=\dot{M_*}/M_*$), as is required based on the comparison with observational data in Figure \ref{fig:ssfrz}.

The break in the GSMF in the simulation evolves in a similar way to that observed, however, between redshifts 2 and 4 there is too little mass in simulated galaxies above $10^{11}$\msun, suggesting that less efficient AGN feedback (or stellar feedback in high mass objects) at high redshifts is required to produce the observed evolution of the break in the GSMF.
Less efficient AGN feedback at high redshifts would also result in more star formation around the peak epoch of star formation, at redshift two, as favoured by current observational data for the star formation rate density.
The requirement for weaker AGN feedback, however, is very sensitive to the stellar mass errors that arise from inferring the GSMF from observations.
While recent observations of the GSMF are typically consistent with each other within their error bars, it is important to consider both random and systematic uncertainties in inferring stellar mass from observed flux, as shown in Figure \ref{fig:masserrors}.
As a result of the sensitivity of the exponential break in the GSMF to the stellar mass errors, it is difficult to determine if the AGN are indeed overly effective in the simulation.

The largest discrepancy we find with observational data is in the SSFRs of star forming galaxies, which are 0.2 to 0.5 dex below the values inferred from observations across all of cosmic time (Figure \ref{fig:ssfrz}).  
This discrepancy cannot be explained as a simple systematic offset in the simulation, as we have shown the stellar mass density to be consistent with observations to within 0.1 dex.
Applying a systematic boost to the star formation rates of 0.3 dex would undo the agreement in the stellar mass density.
It is puzzling that the SSFRs are systematically low, yet the stellar mass growth is consistent with the observational data.
However, we have also found that the galaxy passive fractions appear too low by up to 15 per cent between $10^{10.5}$ and $10^{11.5}$\msun\ (Figure \ref{fig:ssfrevo}).
Assuming that the observed star formation rates are accurate, a potential solution to the low SSFRs is that the star formation is not sufficiently bursty.
More bursty episodes of star formation could produce the same stellar mass with higher star formation rates over shorter time periods than in the current simulation.
This solution has the advantage that it would also increase the passive fractions, as galaxies would be star forming for a smaller fraction of the time.

Observed stellar masses and star formation rates are uncertain at the 0.1 to 0.3 dex level across all observed redshifts.
Until recently hydrodynamical simulations have struggled to reproduce redshift zero galaxy populations within the observational uncertainties, not to mention the evolution of the galaxy population.
The simultaneous comparison to stellar masses and star formation rates across cosmic time thus provides a stringent test for the evolution of galaxy properties in our galaxy formation model.
The \Eagle\ \bigbox\ simulation performs relatively well in this test, verifying that the simulation produces galaxies with reasonable formation histories, for a redshift zero galaxy population that is representative of the observed Universe.
The agreement with observational data from redshifts 0 to 7 is at the level of the systematic uncertainties and follows the observed evolutionary trends.
This gives us confidence that the model can be used as a reliable tool for interpreting observations and to explore the physics of galaxy formation.
To give further confidence, our simulation shows weak numerical convergence, as defined in Section \ref{sec:simres}, of the GSMF to within 0.1 dex for galaxies of stellar masses greater than 100 baryonic particles\footnote{Strong numerical convergence tests are presented in Appendix \ref{ap:res}.} and of the SSFRs to within 0.1 dex when star formation rates are resolved by a minimum of 100 star forming particles when going to a factor of 8 higher resolution.  
This level of convergence enables us to extend the galaxy population to lower stellar masses, by a factor of 8, using \highres, the higher-resolution simulation.

While there is scope to improve agreement with observational data, it is not clear that this should currently be a priority for a number of reasons.
Given that the level of systematic uncertainty in the observations are similar to the level of agreement with the simulation, better agreement with observations would not automatically translate into more confidence in the model.
Secondly, as hydrodynamical simulations are computationally expensive, full parameter space searches are unfeasible using current technology.
Finally, it is likely that achieving better agreement with observations would require more complex parameterisation of the subgrid models, which would be better motivated if changes were supported by small scale simulations modelling ISM physics and smoothed to the resolution of current cosmological simulations.  
While many studies of this kind are underway \citep[e.g.][]{Creasey13}, they do not yet model all the relevant physics and currently require too much computational time to be incorporated into full cosmological simulations.

\section{Summary}
\label{sec:summary}

We have compared the build-up of the stellar mass density, and the evolution of the galaxy stellar mass function and galaxy star formation rates in the \eagle\ cosmological simulations to recent observations. 
The \eagle\ suite includes cosmologically representative volumes of up to (100 cMpc)$^3$, as well as smaller boxes run with higher numerical resolution to assess convergence and to extend the results to lower-mass galaxies. 
The simulations include physically motivated subgrid models for processes that cannot be resolved, with parameters calibrated to reproduce the observed redshift $z\sim 0$ galaxy stellar mass function and galaxy sizes. 
\eagle\ is described in detail and compared with a variety of observations of the present-day Universe in \cite{Schaye14}. 
In this paper we investigated whether the good agreement between simulations and observations of galaxy masses and star formation rates at $z\sim0$ extends to higher redshift, $z=0\rightarrow 7$. 
Our main findings are as follows:

\begin{itemize}
 \item The stellar mass density in the simulation tracks the observed value to within 20 per cent across cosmic time (Figure \ref{fig:smdensity}).
 Observed trends in the evolution of the galaxy stellar mass function are reproduced to within plausible observational uncertainties, over the full redshift range $z=0\rightarrow 7$ (Figure \ref{fig:smfevo}). 
  \item The observed shape of the evolution of the star formation rate density (Figure \ref{fig:sfrh}), and the trends of specific star formation rate, $\dot{\rm  M}_\star/{\rm M}_\star$, as a function of stellar mass and lookback time (Figure \ref{fig:ssfrevo}, \ref{fig:ssfrz}), are all reproduced accurately.
The fraction of passive galaxies increases with stellar mass in the simulation, in agreement with the observed trend (Figure \ref{fig:passive}).
 \item Below stellar masses of $\sim 10^{10.5}$\msun\, the normalisation of the galaxy stellar mass function is above the observations by $\sim$~0.2~dex at redshift 2. 
There is a similar offset in the normalisation of the specific star formation rates, which are low by 0.2-0.5 dex across all redshifts. 
The recent papers of \cite{Mitchell14, White14} highlighted a similar discrepancy with the data, based on semi-analytical models. 
These apparent discrepancies may result from systematic uncertainties in the observations. However, if they are real, then this would imply that even stronger feedback is required at high redshift than what is currently implemented in \eagle. 
Burstier star formation histories could possibly also resolve the apparent discrepancy.
\item Galaxy stellar mass functions and star formation rates are reasonably well converged across all redshifts at which the convergence can be tested (Figure \ref{fig:smfevo}, \ref{fig:ssfrevo}).
\end{itemize}

%A follow-up paper, \citep{Furlong14b}, will show how galaxies of different masses contribute to the evolution of the global statistics presented here, and how the predicted evolution relates to ``galaxy downsizing''.

\section*{Acknowledgements}

	The authors are very grateful for the endless technical support provided by Dr. Lydia
	Heck during the preparation of these simulations and during post processing.
	MF thanks Violeta Gonzalez-Perez and Peter Mitchell for providing observational data sets.

RAC is a Royal Society University Research Fellow.
This work used the DiRAC Data Centric system
at Durham University, operated by the Institute for Computational Cosmology on behalf of the STFC DiRAC HPC
Facility (www.dirac.ac.uk). This equipment was funded by
BIS National E-infrastructure capital grant ST/K00042X/1,
STFC capital grant ST/H008519/1, and STFC DiRAC
Operations grant ST/K003267/1 and Durham University.
DiRAC is part of the National E-Infrastructure. We also
gratefully acknowledge PRACE for awarding us access to
the resource Curie based in France at Tr\'es Grand Centre de Calcul. This work was sponsored by the Dutch National Computing Facilities Foundation (NCF) for the use
of supercomputer facilities, with financial support from the
Netherlands Organization for Scientific Research (NWO).
The research was supported in part by the European Research Council under the European Union's Seventh Framework Programme (FP7/2007-2013) / ERC Grant agreements 278594-GasAroundGalaxies, GA 267291 Cosmiway,
and 321334 dustygal, the Interuniversity Attraction Poles
Programme initiated by the Belgian Science Policy OWNce
([AP P7/08 CHARM]), the National Science Foundation under Grant No. NSF PHY11-25915, the UK Science and Technology Facilities Council (grant numbers ST/F001166/1 and
ST/I000976/1), Rolling and Consolidating Grants to the
ICC, Marie Curie Reintegration Grant PERG06-GA-2009-256573, Marie Curie Initial Training Network Cosmocomp (PITN-GA-2009-238356).

%Add bibliography
%\bibliographystyle{plainnat} 
\bibliographystyle{mn2efix} 
\bibliography{../../mybib}

%Appendicies
\appendix
\section{Schechter function fits}
\label{ap:Sch}

\begin{table*} 
\centering  
\caption{Single (eq. \ref{eq:Sch}) and double (eq. \ref{eq:Sch_d}) Schechter function parameters for the \Eagle\ \bigbox\ GSMFs presented in Figure \ref{fig:smfevo}, fitting over the mass range $10^8$ to $10^{12}$\msun.
One sigma errors, determined from the covariance matrix, are also listed.
The Schechter function parameters provide a simple way of reproducing the GSMFs from the \Eagle\ simulation over the range where the fitting is carried out.
} 
\smallskip 
 \begin{minipage}{17cm} 
 \centering 
  \begin{tabular}{|c |c | c| c| c| c|} 
    \hline
	{\bf Redshift} & {\bf log$_{10}$(M$_C$)} & {\bf $\Phi^*$} & {\bf $\alpha$} &  & \\
    \hline
     & $\lbrack$M$_\odot \rbrack$ &  {$\lbrack$ $10^{-3}$ cMpc$^{-3} \rbrack$} & &  & \\
	\hline
0.1 & 11.14 $\pm$ 0.09 & 0.84 $\pm$ 0.13 & -1.43 $\pm$ 0.01 & - & - \\
0.5 & 11.11 $\pm$ 0.08 & 0.84 $\pm$ 0.12 & -1.45 $\pm$ 0.01 & - & - \\
1.0 & 11.06 $\pm$ 0.08 & 0.74 $\pm$ 0.10 & -1.48 $\pm$ 0.01 & - & - \\
2.0 & 10.91 $\pm$ 0.08 & 0.45 $\pm$ 0.07 & -1.57 $\pm$ 0.01 & - & - \\
3.0 & 10.78 $\pm$ 0.11 & 0.22 $\pm$ 0.05 & -1.66 $\pm$ 0.01 & - & - \\
4.0 & 10.60 $\pm$ 0.15 & 0.12 $\pm$ 0.04 & -1.74 $\pm$ 0.02 & - & - \\
	\hline
	{\bf Redshift} & {\bf log$_{10}$(M$_C$)} & {\bf $\Phi^*_1$} & {\bf $\alpha_1$} & {\bf $\Phi^*_2$} & {\bf $\alpha_2$}\\
    \hline
     & $\lbrack$M$_\odot \rbrack$ &  {$\lbrack$ $10^{-3}$ cMpc$^{-3} \rbrack$} & & {$\lbrack$ $10^{-3}$ cMpc$^{-3} \rbrack$} & \\
	\hline
0.1 & 10.95 $\pm$ 0.03 & 1.45 $\pm$ 0.12 & -1.31 $\pm$ 0.03 & 0.00 $\pm$ 0.00 & -2.22 $\pm$ 0.22 \\
0.5 & 10.88 $\pm$ 0.04 & 1.61 $\pm$ 0.13 & -1.24 $\pm$ 0.08 & 0.08 $\pm$ 0.10 & -1.79 $\pm$ 0.15 \\
1.0 & 10.74 $\pm$ 0.05 & 1.51 $\pm$ 0.18 & -0.98 $\pm$ 0.17 & 0.48 $\pm$ 0.20 & -1.62 $\pm$ 0.05 \\ 
2.0 & 10.44 $\pm$ 0.08 & 1.06 $\pm$ 0.24 & -0.25 $\pm$ 0.29 & 0.80 $\pm$ 0.14 & -1.58 $\pm$ 0.02 \\
3.0 & 10.19 $\pm$ 0.09 & 0.63 $\pm$ 0.13 & 0.23 $\pm$ 0.37 & 0.61 $\pm$ 0.13 & -1.64 $\pm$ 0.02 \\
4.0 & 10.00 $\pm$ 0.11 & 0.24 $\pm$ 0.08 & 0.89 $\pm$ 0.58 & 0.43 $\pm$ 0.12 & -1.69 $\pm$ 0.03 \\
	\hline
  \end{tabular}  \par
 \end{minipage}
 \label{table:Schechter}
\end{table*}

\begin{figure*}
  \centering
  \includegraphics[width=1.0\textwidth]{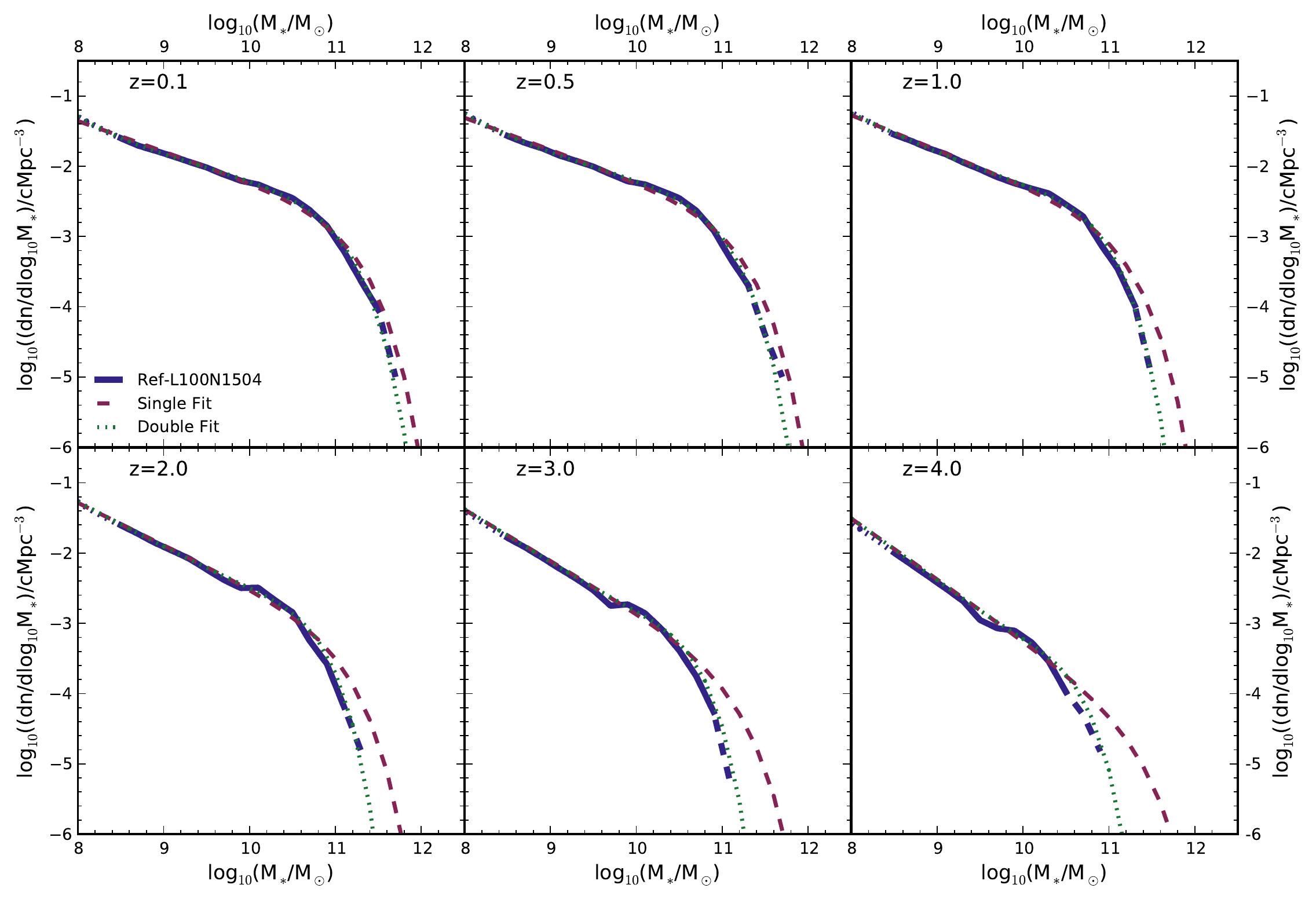}
  \caption{ 
The GSMF from the \bigbox\ simulation (blue), using single Schechter fits (red dashed) and using double Schechter fits (green dotted) at 6 redshifts.
The parameters for the fitting functions can be found in Table \ref{table:Schechter}.
	}
  \label{fig:SchFits}
\end{figure*}

\begin{figure}
  \centering
  \includegraphics[width=0.5\textwidth]{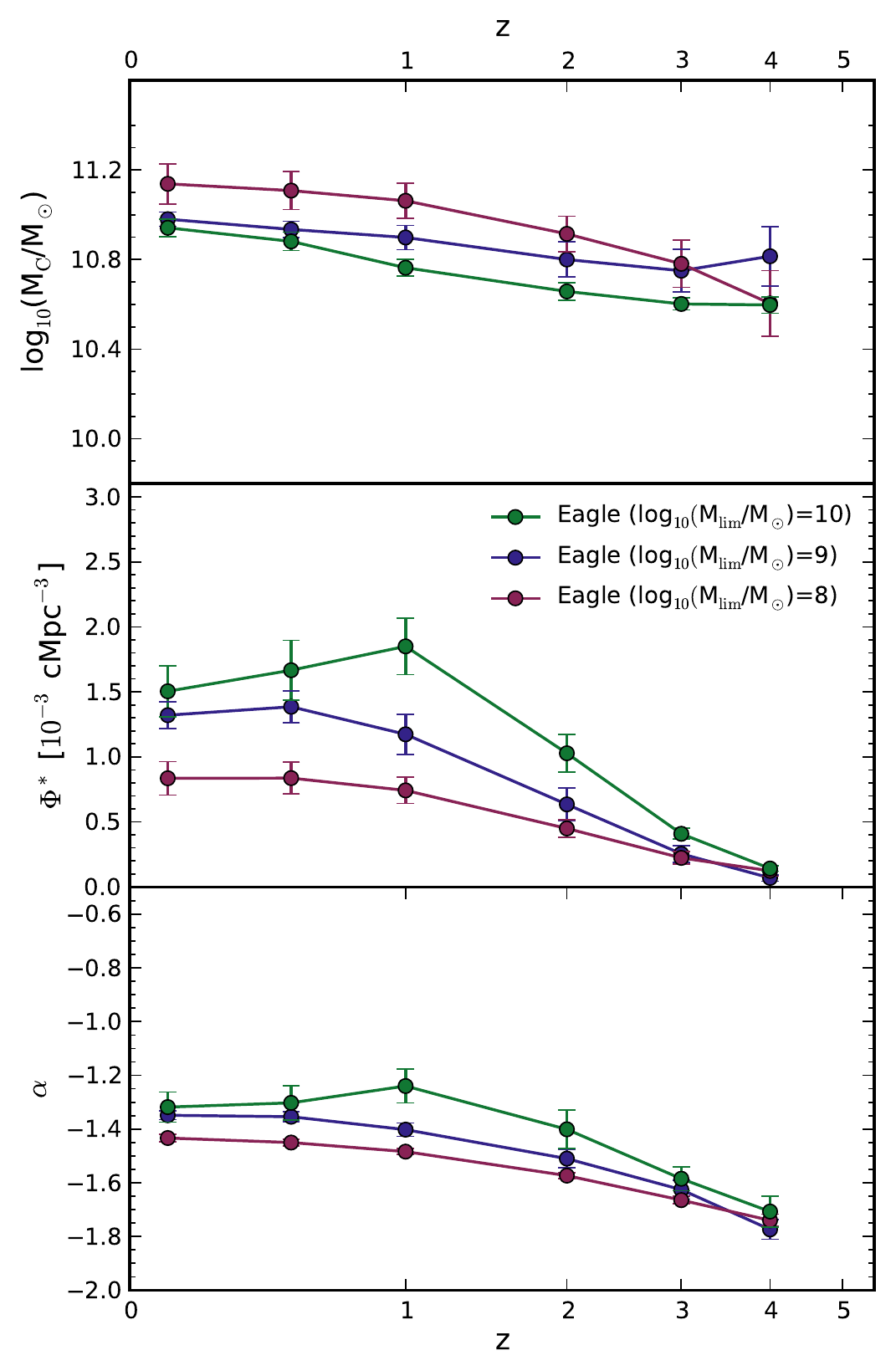}
  \caption{ 
	The Schechter function parameters, \Msch, \Phis\ and \alphas\  for the \Eagle\ GSMFs (as shown in Figure \ref{fig:smfevo}) as a function of redshift.
	These panels show single Schechter function parameters fit from $10^8$, $10^9$ and $10^{10}$\msun\ to $10^{12}$\msun\ in red, blue and green respectively, with 1-$\sigma$ error bars from the fitting.
	The Schechter function fitting is sensitive to the mass range over which the fitting is done and the values for both \Msch\ and \Phis\ are degenerate.
	For double Schechter function parameters the agreement between different stellar mass ranges is worse due to the increased freedom (not shown).
	}
  \label{fig:SchParams}
\end{figure}

To provide a simple way of reproducing the \Eagle\ GSMFs and to quantify the trends seen in the evolution of the normalisation and the exponential break, we have fit the \Eagle\ GSMFs with Schechter functions.
We fit the GSMFs of \bigbox\ from redshifts 0.1 to 4 that were shown in Figure \ref{fig:smfevo} (blue curves) with single Schechter functions (eq. \ref{eq:Sch}) and double Schechter functions,
\begin{align}
\Phi(M) dM =& \left[ \Phi^*_1 \left( \frac{M}{M_C}\right)^{\alpha_1}  + \Phi^*_2 \left(\frac{M}{M_C}\right)^{\alpha_2} \right] e^{-M/M_C} dM ,
\label{eq:Sch_d}
\end{align}
which is the sum of two Schechter functions with the same characteristic mass, \Msch, but different normalisations, \Phione\ and \Phid\ and different low-mass slopes, \alphaone\ and \alphad.
Double Schechter fits are increasingly used in observational studies fitting the GSMF.
We use least squares fitting with bins of width 0.2 dex in stellar mass.
Bins are weighted by their Poisson error, thereby down weighting the poorly sampled galaxies in the most massive stellar mass bins.
The fits over the mass range $10^8$ to $10^{12}$\msun\ are presented in Table \ref{table:Schechter}.
These fits compared to the simulation data can be seen in Fig. \ref{fig:SchFits}.

To understand the dependence of the Schechter function parameters on the fitted mass range, we applied our fitting routine over three mass ranges, from $10^8$, $10^9$ and $10^{10}$ to $10^{12}$\msun.
Figure \ref{fig:SchParams} shows the evolution of the Schechter function parameters \Msch, \Phis\ and \alphas\ for the single Schechter function fits.
For the single Schechter fit \Msch\ drops over the redshift range zero to four for all mass ranges.
However, the extent of the decrease depends on the fitting range. 
For example, there is a decrease of 0.5 dex when fitting above $10^8$\msun\ compared to a 0.3 dex decrease for fits above $10^9$ and $10^{10}$\msun.  
\Phis\ is reasonably flat until redshift one, with a decrease at redshifts above one for all fits.
There is however an obvious difference in the value of \Phis\ recovered for different fitting ranges, and there is also a difference in their variation with redshift.
The opposite changes in \Msch\ and \Phis\ for the different mass ranges highlight the degeneracy between these two parameters.

The \alphas\ parameter becomes more negative with increasing redshift for fits above $10^{8}$ and $10^9$\msun, showing that the low mass slope steepens with redshift.
However, different behaviour is seen for fits above $10^{10}$\msun\ where \alphas\ increases to redshift 1, then decreases.  
This is not unexpected given that fitting for stellar masses above $10^{10}$\msun\ does not provide enough information to constrain the slope for masses $\ll$\Msch.

We find larger differences between different mass ranges, and in particular larger error bars, when fitting double Schechter functions than what is presented for single Schechter functions in Figure \ref{fig:SchParams}.
Due to the sensitivity of the Schechter fitting to the mass range over which it is done, it is very difficult to compare the fitting parameters directly to observations.
This is especially true when we consider the evolving mass completeness limit for observations.  Any trends with redshift could easily be a result of the changing mass range.
The degeneracy between \Msch\ and \Phis\ also makes a comparison of Schechter parameters difficult to interpret.
The final issue with directly comparing Schechter parameters from observations and/or simulations is the sensitivity of the break in the Schechter function to stellar mass errors, as shown in Section \ref{sec:errors}.
As a result of these issues, we choose not to compare the Schechter function parameters to those determined observationally and consider the comparison of the data presented in Figure \ref{fig:smfevo}, from which Schechter parameters are derived, to be sufficient to determine the agreement between observations and simulations.
However, the Schechter function parameter do provide a simple way of representing the GSMFs from the Eagle simulation over the range where the fitting is carried out.

\section{Strong numerical convergence}
\label{ap:res}

\begin{figure}
  \centering
 \begin{minipage}{0.48 \textwidth}
  \includegraphics[width=1.0\textwidth]{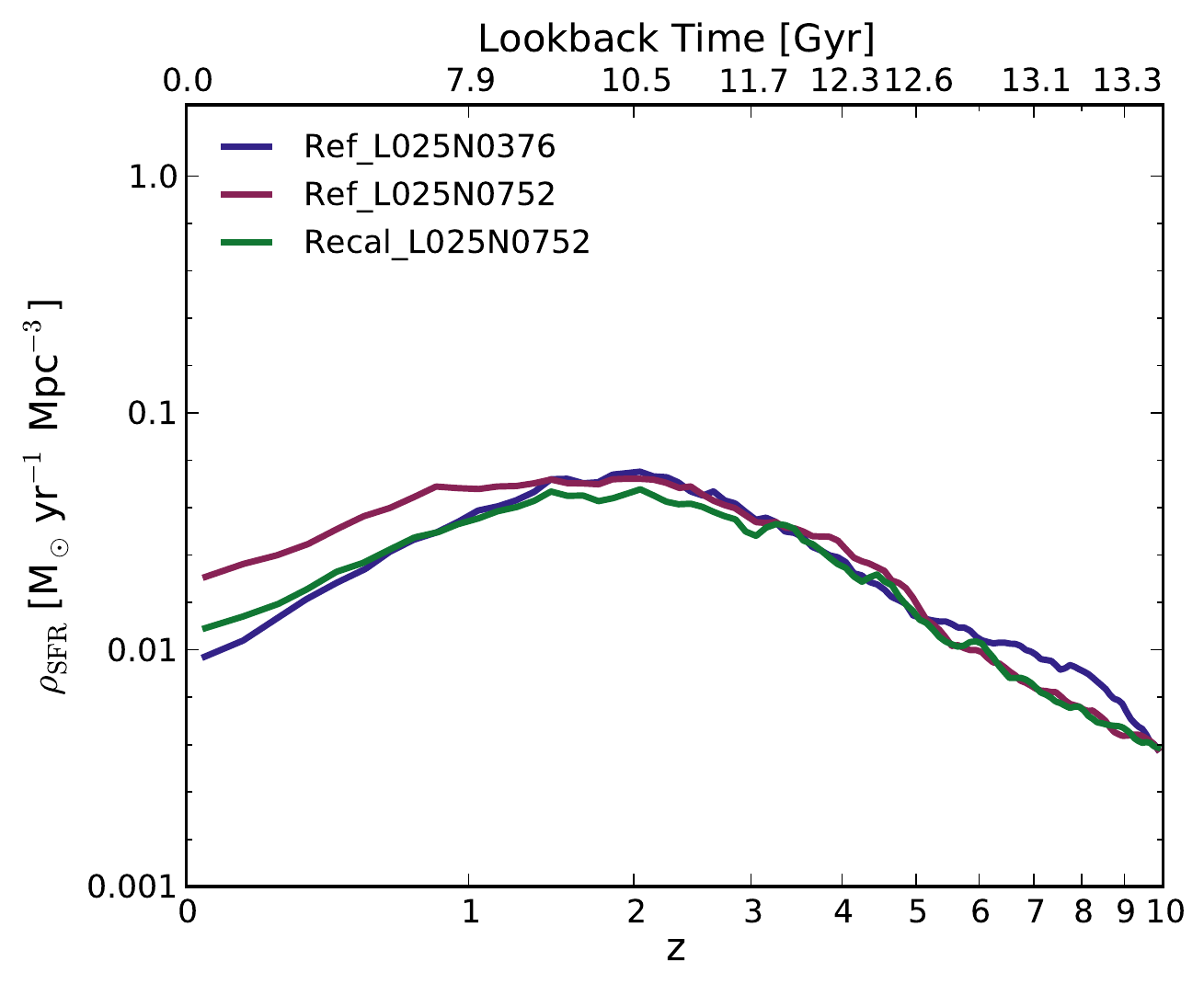}
 \end{minipage} \hfill
 \begin{minipage}{0.48 \textwidth}
  \includegraphics[width=1.0\textwidth]{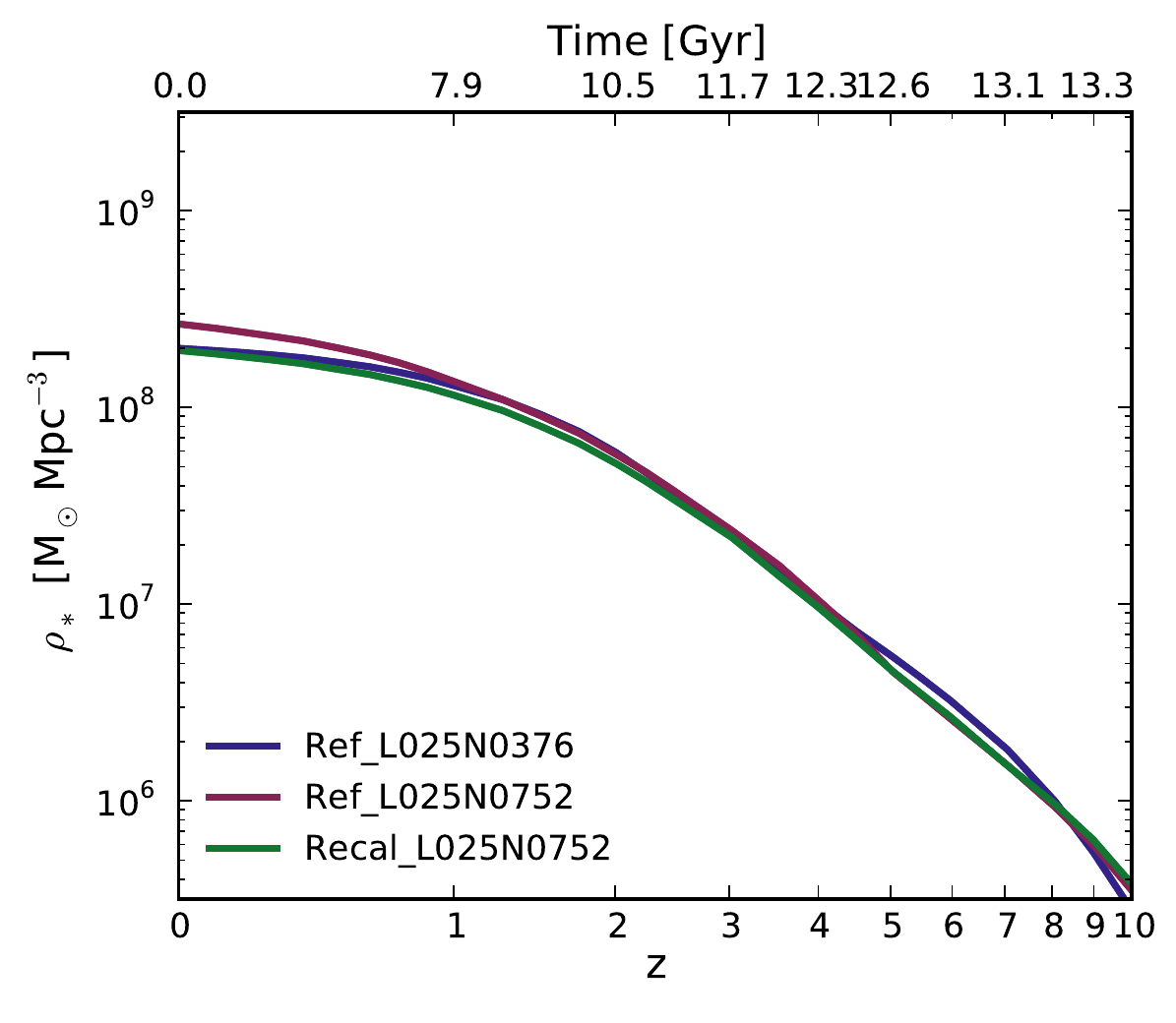}
 \end{minipage} \hfill
  \caption{
The star formation rate density and stellar mass density as a function of redshift in the top and bottom panels for \smallbox, using the same physics as for \bigbox\ shown in all previous plots, Ref-L025N0752, with eight times the resolution and \highres, with eight times higher resolution and recalibrated subgrid parameters in blue, red and green respectively.
	}
  \label{fig:res_all}
\end{figure}

\begin{figure*}
  \centering
  \includegraphics[width=1.0\textwidth]{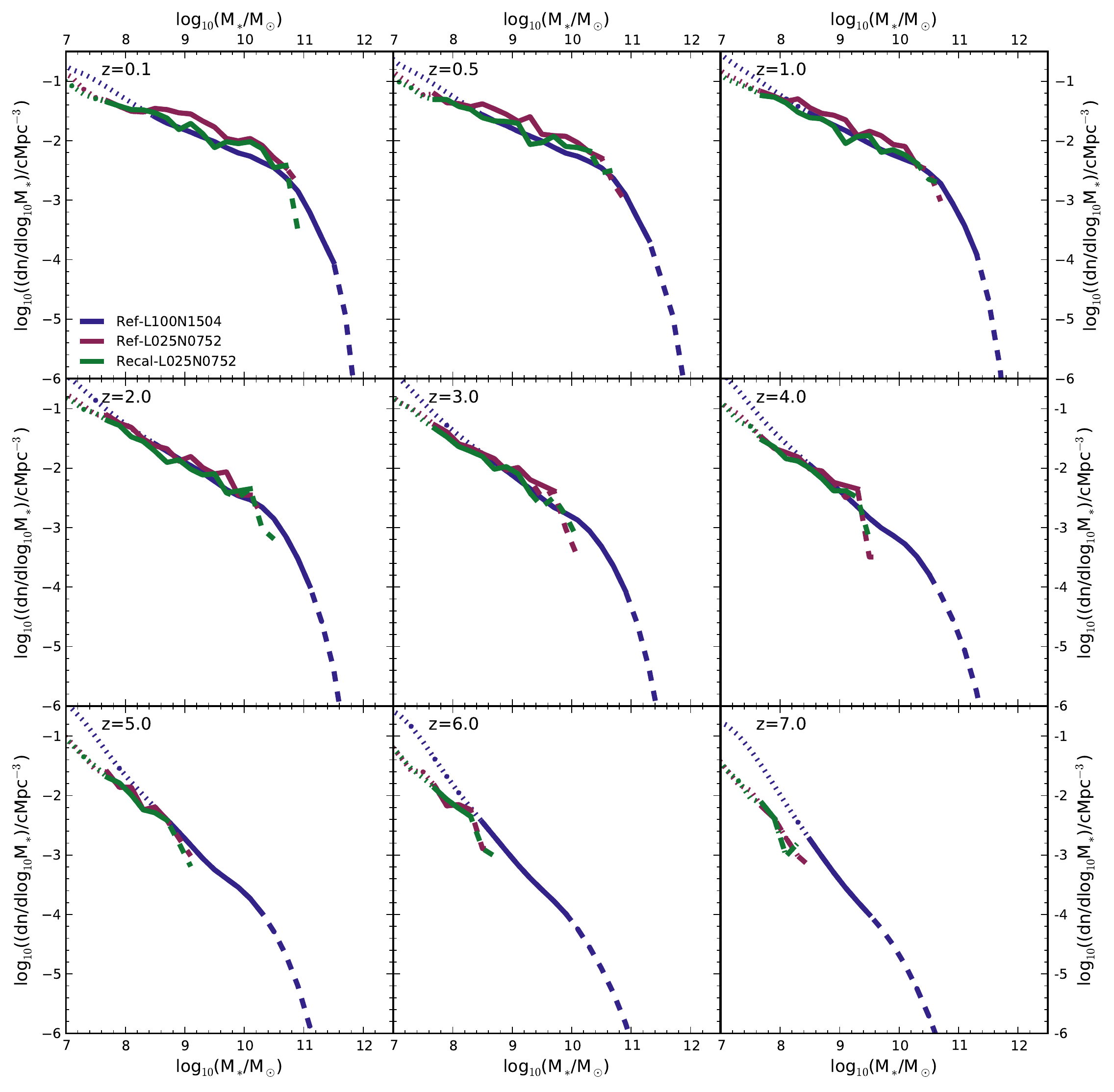}
  \caption{ 
The GSMF from \bigbox\ (blue), \highresref\ (red) and \highres\ (green) at 9 redshifts.
At redshift 0.1 the \highres\ simulation is within $0.1$ dex of \bigbox, while the \highresref\ simulation is within $0.2$ dex.
The agreement between intermediate and high resolution simulations improves with increasing redshift.
	}
  \label{fig:res_smf}
\end{figure*}

Here we show strong and weak resolution tests for the evolution of the global stellar mass and star formation rate densities in Fig. \ref{fig:res_all}.
Three models are considered, \smallbox, equivalent in resolution and model parameters to \bigbox\ except in a 25 cMpc box as opposed to 100 cMpc; Ref-L025N0752, with the same subgrid parameters as \bigbox\ but with 8 times higher mass resolution in a 25 cMpc box; and \highres, with recalibrated subgrid parameters and 8 times higher mass resolution than \bigbox\ in a 25 cMpc box.
The 25 cMpc boxes for which we have higher-resolution simulations are too small to be representative.
To ensure we do not obscure the effects of resolution with other effects such as box size, we compare the same box size for all models.

Figure \ref{fig:res_all} shows \sfrd\ for all three 25 cMpc simulations in the top panel.
Between redshifts 9 and 5 the \smallbox\ simulation has an excess of star formation relative to both higher-resolution simulations, of less than 0.2 dex, which results from the coarser minimum star formation rate per particle at the standard resolution.
The largest difference between the 3 simulations is at redshift 0.1, where the \highresref\ has a higher \sfrd\ by 0.3 dex.
The \smd\ is shown in the bottom panel of Figure \ref{fig:res_all}. 
As \smd\ is the integral of \sfrd\, modulo stellar mass loss, the differences seen here, at redshifts above 4 for \smallbox\ and at redshift zero for \highresref\, reflect those seen in \sfrd.

In Fig. \ref{fig:res_smf} again three models are compared, in this case \bigbox, \highres\ and \highresref.
The agreement between \bigbox\ and \highres, testing weak convergence, is around 0.1 dex at redshift 0.1 over the range of stellar masses that can be probed, as reported in Section \ref{sec:weakcon}. 
The agreement is similar across all redshift ranges.
Comparing \bigbox\ and \highresref, to test strong convergence, the stellar mass functions agree to within $\sim 0.2$ dex at redshift 0.1 and the agreement improves with increasing redshift.  
At redshifts 4 and above the level of agreement is similar to \highres.
In \Schaye\ the redshift 0.1 strong convergence was found to be similar to that obtained by simulations from other groups \citep[e.g.][]{Vogelsberger13}, while the agreement in \Eagle\ improves at higher redshifts.

Overall the level of agreement shown for the strong, and particularly for the weak convergence, is good.

\bsp

\label{lastpage}

\end{document}